\pdfoutput=1%
\documentclass[a4paper,british,floatfix,onecolumn,pdftex,superscriptaddress,twoside,%
aip,jcp,%
citeautoscript, 
preprint
]{revtex4-2}%
\usepackage{amsfonts,amsmath,amssymb}
\usepackage{commath}%
\usepackage[utf8]{inputenc}
\usepackage[T1]{fontenc}
\usepackage{graphicx}%
\usepackage{microtype}
\usepackage{lmodern}
\usepackage[obeyFinal,textsize=footnotesize]{todonotes}
\usepackage{xspace}
\usepackage{hyperref, hypernat}
\usepackage[todos]{CMImacros}

\fxsetup{final}
\hypersetup{citebordercolor=yellow,linkbordercolor=red,urlbordercolor=blue}

\newcommand{\cfeldesy}{\affiliation{Center for Free-Electron Laser Science, Deutsches
      Elektronen-Synchrotron DESY, Notkestraße 85, 22607 Hamburg, Germany}}%
\newcommand{\uhhcui}{\affiliation{Center for Ultrafast Imaging, Universität Hamburg, Luruper
      Chaussee 149, 22761 Hamburg, Germany}}%
\newcommand{\uhhphys}{\affiliation{Department of Physics, Universität Hamburg, Luruper Chaussee 149,
      22761 Hamburg, Germany}}%
\newcommand{\mbi}{\affiliation{Max Born Institute, Max-Born-Straße 2a, 12489 Berlin, Germany}}%
\newcommand{\jkemail}{\email[]{jochen.kuepper@cfel.de}}%
\newcommand{\aremail}{\email[]{arnaud.rouzee@mbi-berlin.de}}%
\newcommand{\cmiweb}{\homepage{https://www.controlled-molecule-imaging.org}}%

\begin{document}
\title{Time-resolving the UV-initiated photodissociation dynamics of OCS}%
\author{Evangelos T.\ Karamatskos}\cfeldesy\uhhphys%
\author{Suresh Yarlagadda}\mbi%
\author{Serguei Patchkovskii}\mbi%
\author{Marc~J.~J.~Vrakking}\mbi%
\author{Ralph~Welsch}\cfeldesy%
\author{Jochen~Küpper}\jkemail\cmiweb\cfeldesy\uhhphys\uhhcui%
\author{Arnaud~Rouz\'{e}e}\aremail\mbi%
\date{\today}
\begin{abstract}
   We present a time-resolved study of the photodissociation dynamics of OCS after
   UV-photoexcitation at $\lambda=237$~nm. OCS molecules ($X\,^1\Sigma^+$) were primarily excited to
   the $1\,^1\!A''$ and the $2\,^1\!A'$ Renner-Teller components of the $^1\Sigma^{-}$ and
   $^1\!\Delta$ states. Dissociation into CO and S fragments was observed through time-delayed
   strong-field ionisation and imaging of the kinetic energy of the resulting CO$^+$ and S$^+$
   fragments by intense 790~nm laser pulses. Surprisingly, fast oscillations with a period of
   $\ordsim100$ fs were observed in the S$^+$ channel of the UV dissociation. Based on
   wavepacket-dynamics simulations coupled with a simple electrostatic-interaction model, these
   oscillations do not correspond to the known highly-excited rotational motion of the leaving
   CO$(X\,^1\Sigma^+,J\gg0)$ fragments, which has a timescale of $\ordsim140$~fs. Instead, we
   suggest to assign the observed oscillations to the excitation of vibrational wavepackets in the
   $2\,^3\!A''$ or $2\,^1\!A''$ states of the molecule that predissociate to form S$(^3\!P_{J})$
   photoproducts.
\end{abstract}
\maketitle

\section{Introduction}
\label{sec:intro}
Imaging ultrafast photochemical reactions with quantum-mechanical detail and atomic temporal and
spatial resolution is one of the ultimate dreams of current physical chemistry and the molecular
sciences in general~\cite{Zewail:JPCA104:5660, Young:JPB51:032003, Barty:ARPC64:415}. Various
approaches have been discussed, ranging from x-ray~\cite{Barty:ARPC64:415, Pande:Science352:725,
   Kuepper:PRL112:083002, Stern:FD171:393} and electron diffractive imaging~\cite{Zewail:ARPC57:65,
   Ischenko:CR117:11066} to laser-induced electron diffraction(LIED)~\cite{Blaga:Nature483:194,
   Wolter:Science354:308} and Coulomb-explosion ion imaging~\cite{Stapelfeldt:PRL74:3780,
   Pitzer:Science341:1096, Nagaya:FD194:537}. Complementary methods such as high-harmonic
spectroscopy~\cite{Itatani:Nature432:867, Vozzi:NatPhys7:822, Woerner:Nature466:604} and
time-resolved photoelectron spectroscopy~\cite{Baumert:PRL64:733, Stolow:ACP139:497, Boll:FD171:57}
have provided further insight into these processes. While the experimental recording of the
envisioned high-resolution ``molecular movies'' are still outstanding, promising examples
demonstrate the progress that has been made, for instance, using time-resolved electron
diffraction~\cite{Williamson:Nature386:159, Yang:Science361:64, Wolf:NatChem11:504}. We are
interested in exploring the possibilities for atomically resolved imaging of photoinduced chemical
reactions using LIED~\cite{Karamatskos:JCP150:244301, Trabattoni:NatComm11:2546,
   Karamatskos:thesis:2019}. In this paper, we present ion-imaging measurements that we have
performed as the precursor to such a LIED experiment.

The carbonyl sulfide (OCS) molecule provides an interesting benchmark system and serves as an
important testbed for the study of photodissociation dynamics. OCS is linear in its rovibronic
ground state $(X\,^1\Sigma^{+})$, but it bends in its first absorption band upon photoexcitation in
the range from $190$~nm to $255$~nm~\cite{Rabalais:CR71:73}. Simultaneously to bending, the OCS
molecule dissociates into S and CO. An advantage of studying the photodissociation dynamics of OCS
lies in the fact that being an asymmetric triatomic molecule, symmetric and antisymmetric stretching
vibrations can be uncoupled, and since the CO fragment is produced mainly in its electronic and
vibrational ground state CO$(X\,^1\Sigma^{+},v=0)$~\cite{Sivakumar:JCP88:3692, Sato:JPC99:16307},
the stretching motion of the stiff CO bond can be considered a spectator during the dissociation.
This significantly simplifies the understanding of the dissociation dynamics and allows for the use
of a reduced dimensionality model~\cite{Suzuki:JCP109:5778}. The photodissociation dynamics of OCS
in its first absorption band has been extensively studied, both
experimentally~\cite{Breckenridge:JCP52:1713, Sivakumar:JPC89:3609, Sivakumar:JCP88:3692,
   Katayanagi:CPL247:571, Sato:JPC99:16307, Suzuki:JCP109:5778,
   Kim:JPCA103:10144,Sugita:JCP112:7095, Rakitzis:JCP111:10415, Katayanagi:CPL360:104,
   Rijs:JCP116:2776, Wei:JCP145:024310, Brouard:JCP127:084304, VanDenBrom:JCP121:11645,
   VanDenBrom:JCP123:164313, Rakitzis:Science303:1852, Lipciuc:PCCP8:3007, Lipciuc:PCCP13:8549,
   Sofikitis:PRL118:253001, Sofikitis:PRA98:033417} and theoretically~\cite{Suzuki:JCP109:5778,
   Schmidt:JCP137:054313, Schmidt:ACP13:1511, Schmidt:JCP136:131101, McBane:JCP138:094313,
   Danielache:JCP131:024307}. Also the atmospheric relevance of OCS and its photodissociation as a
sulphur source are of ongoing interest~\cite{Buehl:AtmosChemPhys12:1239, Hattori:PNAS117:20447}.

For linear ground-state OCS dipole transitions to the excited $^1\Sigma^{-}$ and $^1\!\Delta$ states
are forbidden by symmetry. However, upon bending due to Renner-Teller interactions, degeneracies are
lifted and the transitions become weakly allowed, with the transition strenghts increasing with
increasing bending angle~\cite{Lipciuc:JCP126:194318}. In its first absorption band, OCS dissociates
predominantly into
\begin{align}
   \text{OCS} + h\nu &\longrightarrow \text{CO}(X\,^1\Sigma^{+}) + \text{S}(^1\!D_{2}) \\
   \intertext{and}
   \text{OCS} + h\nu &\longrightarrow \text{CO}(X\,^1\Sigma^{+}) + \text{S}(^3\!P_{J}), \quad J=0,1,2
\end{align}
with a branching ratio of $95:5$ at the center of the absorption band at
222~nm~\cite{Suzuki:JCP109:5778}. In both channels, the CO fragment is mainly produced in its
electronic and vibrational ground state. An upper bound for the population in the first excited
vibrational state was given to be about 2~\%~\cite{Sivakumar:JCP88:3692, Sato:JPC99:16307}. The CO
fragment produced in the major dissociation channel, together with a $\text{S}(^1\!D_{2})$ fragment,
is highly rotationally excited, showing a bimodal rotational state distribution with two
Gaussian-shaped underlying distributions. These distributions peak around $J=55$ and $J=67$ for
photodissociation at $222$~nm~\cite{Sivakumar:JPC89:3609, Sivakumar:JCP88:3692,
   Katayanagi:CPL247:571} and shift towards lower angular momenta for longer ultraviolet
(UV)-excitation wavelength~\cite{Sivakumar:JCP88:3692, Suzuki:JCP109:5778}. The origin of the
bimodal distribution was attributed to two different dissociation scenarios: The small-$J$-state
component, measured at higher product kinetic energies, is generally assigned to direct dissociation
from the $2\,^1\!A'$ ($1\,^1\!\Delta$) and $1\,^1\!A''$ ($1\,^1\Sigma^{-}$) states. The
large-$J$-state component observed at low kinetic energies is due to the dissociation from the
$2\,^1A'$ ($1\,^1\!\Delta$) state alone, but involving non-adiabatic transitions to the $1\,^1A'$
($X\,^1\Sigma^{+}$) ground state~\cite{Suzuki:JCP109:5778}. For graphical representations of the
energies of all electronic states involved we refer the reader to, \eg, \onlinecite[Figures~1 in
refs.][]{Schmidt:JCP136:131101, Toulson:JPCA120:6745}.

Although the dissociation process responsible for the formation of rotationally excited
CO($X\,^1\Sigma^{+}$) and $\text{S}(^1\!D_{2})$ products is well understood and mainly attributed to
prompt dissociation following the excitation of the $2\,^1\!A'$ ($1\,^1\!\Delta$) and $1\,^1\!A''$
($1\,^1\Sigma^{-}$) states, the mechanisms responsible for the formation of $\text{S}(^3\!P_{J})$
fragments remains to be elucidated. Strong vibrational resonances were recently observed in the 2+1
resonance-enhanced multi-photon ionisation (REMPI) spectra of the S$(^3\!P_{J})$ photoproducts
recorded in a wavelength range from 212~nm to 260~nm, which were assigned to vibrational
progressions in the C--S stretching following direct excitation to the $2\,^1\!A''$
($1\,^1\!\Delta$) and triplet $2\,^3\!A''$ ($1\,^3\Sigma^{-}$) metastable
states~\cite{Toulson:JPCA120:6745}. Molecules in the $2\,^3\!A''$ ($1\,^3\Sigma^{-}$) state can
rapidly predissociate through non-adiabatic transitions to the repulsive $1\,^3\!A''$ state, leading
to S$(^3P_J)$ products. Alternatively, it has been predicted that the $2\,^1\!A''$ ($1\,^1\!\Delta$)
and $2\,^3\!A''$ ($1\,^3\Sigma^{-}$) states can predissociate through non-adiabatic coupling to the
$1\,^1\!A''$ ($1\,^{1}\Sigma^{-}$) state or through spin-orbit coupling to the $2\,^1\!A'$
($1\,^1\!\Delta$) state, respectively, with subsequent intersystem crossing leading to the formation
of S$(^3\!P_{J})$ fragments~\cite{Toulson:JPCA120:6745}.

So far, the experimental studies of the photodissociation of OCS were limited to frequency-resolved
spectroscopies, \ie, with nanosecond lasers. This includes laser-induced fluorescence (LIF)
spectroscopy~\cite{Sivakumar:JPC89:3609,Sivakumar:JCP88:3692,Nan:CPL209:383} and
REMPI~\cite{Sato:JPC99:16307, Katayanagi:CPL247:571, Mo:PRL77:830, Suzuki:JCP109:5778,
   Kim:JPCA103:10144, Sugita:JCP112:7095, Rakitzis:PRL87:123001, Katayanagi:CPL360:104,
   Rijs:JCP116:2776, Rakitzis:Science303:1852, VanDenBrom:JCP121:11645, VanDenBrom:JCP123:164313,
   Lipciuc:PCCP8:3007, Lee:JCP125:144318, Brouard:JCP127:084304, Lipciuc:PCCP13:8549,
   Wei:JCP145:024310, Sofikitis:PRL118:253001, Sofikitis:PRA98:033417}. These experiments, coupled
to computational investigations~\cite{Schmidt:JCP137:054313, Schmidt:JCP136:131101,
   Schmidt:ACP13:1511, Schmidt:JCP141:184310, McBane:JCP138:094313}, have provided a wealth of
detailed information on the photodissociation products, including their final state distributions.
However, to date no time-resolved studies have been carried out to probe the transient electronic
and nuclear structure of the molecule along the dissociation pathway. Starting with the present
work, we try to bridge this gap. Here, the dissociation dynamics of the OCS molecule following UV
excitation with a $237$~nm femtosecond laser pulse is investigated. Reactants and products are
probed by strong-field ionisation with a time-delayed $790$~nm femtosecond laser pulse and the
velocity distributions of S$^+$ and CO$^+$ fragment ions are imaged. Furthermore, quantum chemistry
simulations taking into account the prompt dissociation following excitation of the $2\,^1\!A'$
($1\,^1\!\Delta$) and $1\,^1\!A''$ ($1\,^1\Sigma^{-}$) states to form S$(^1\!D_{2})$ products were
coupled to a simple model for calculating strong-field-ionisation rates. Interestingly, this model
was unable to reproduce the fast oscillations observed in the kinetic energy distribution of the
S$^+$ fragments. Instead, our results provide strong indications that vibrational wavepackets are
formed in the $2\,^1\!A''$ ($1\,^1\!\Delta$) and $2\,^3\!A''$ ($1\,^3\Sigma^{-}$) states, which
predissociate to form $\text{S}(^3\!P_{J})$. These experiments form the basis for ongoing
investigations of the ultrafast photodissociation dynamics of OCS following UV excitation using
structurally~\cite{Karamatskos:JCP150:244301, Trabattoni:NatComm11:2546, Karamatskos:NatComm10:3364}
or electronic-state~\cite{Wiese:PRR2020:inprep} resolved laser-induced-electron diffraction and
photoelectron imaging experiments.

\section{Experimental setup}
\label{sec:setup}
The experimental setup was described previously~\cite{Karamatskos:NatComm10:3364}. Briefly, a cold
molecular beam was formed by supersonic expansion of a 90~bar mixture of OCS (500~ppm) seeded in
helium using a pulsed Even-Lavie valve~\cite{Hillenkamp:JCP118:8699} operated at 250~Hz. Thus, all
molecules were in their electronic ground state, vibrationally cold, and their rotational
temperature was characterised to be $T_{\text{rot}}=0.6$~K at the interaction point of the molecules
with the laser pulses~\cite{Karamatskos:NatComm10:3364}.

The output of a commercial Ti:Sapphire amplified laser system (KMLabs, Wyvern30), delivering 30~mJ,
38~fs (FWHM) pulses at a central wavelength of 790~nm and a repetition rate of 1~kHz, was divided
into three beams, whose time delays were controlled using motorised delay stages. Up to 20~mJ were
used to pump an optical parametric amplifier (Light Conversion HE-TOPAS), generating ultrashort
pulses at 1190~nm with a maximum output energy of 3.5~mJ. A second path was used to generate UV
light \emph{via} second harmonic generation followed by frequency mixing with the second harmonic of
the 1190~nm laser pulse. This yielded UV pulses with a wavelength of $237$~nm, a pulse duration of
$70$~fs (FWHM), and a pulse energy of up to 15~\uJ. The third 790~nm (NIR) beam was used as the
ionising probe pulse.

The UV-pump and NIR-probe pulses were temporally and spatially overlapped at the centre of a
high-energy velocity-map-imaging spectrometer (VMI) with their polarisation parallel to the plane of
the detector. CO$^{+}$ and S$^{+}$ ions were generated through strong-field ionisation of OCS or its
dissociation products, accelerated into a 10~cm long field-free flight tube, and detected with the
combination of a 77~mm diameter dual-microchannel-plate/phosphor-screen assembly and a CCD camera.
The projected two-dimensional (2D) ion-momentum distributions were inverted using an Abel-inversion
procedure based on the BASEX algorithm~\cite{Dribinski:RSI73:2634} in order to yield
three-dimensional (3D) ion-momentum distributions.

Ion momentum distributions were recorded for OCS$^+$, S$^+$, and CO$^+$ fragments for a series of
pump-probe time delays $\tau$ separated by 20~fs. Cross-correlation measurements of the parent ion
were used to define the time overlap between the UV-pump and the NIR-probe pulse, \ie, $\tau = 0$.
The S$^+$ and CO$^+$ ion momentum distributions were recorded for various sets of UV (4.5~\uJ and
10~\uJ) and NIR (300~\uJ and 1.3~mJ) pulse energies. All ion momentum distributions presented in
this manuscript were averaged over, at least, 3500 laser shots.

\section{Computational wavepacket dynamics}
\label{sec:computation}
\begin{figure}
   \centering%
   \includegraphics[width=0.25\linewidth,angle=0]{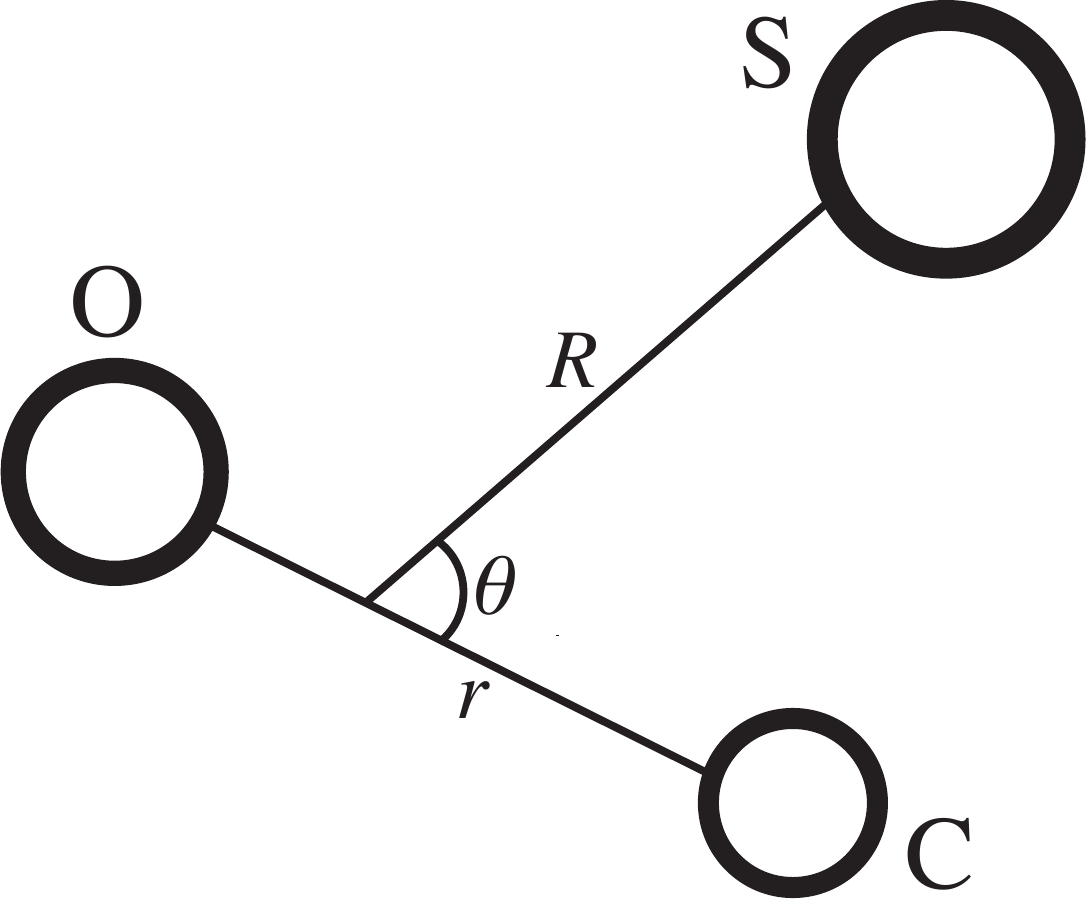}%
   \caption{Jacobi coordinates employed in the quantum dynamics simulations.}
   \label{fig:coordinates}
\end{figure}

The photodissociation of OCS in its first absorption band is known to mainly proceed by prompt
dissociation following the excitation of the $2\,^1\!A'$ ($1\,^1\!\Delta$) and $1\,^1\!A''$
($1\,^1\Sigma^{-}$) states. Therefore, in our calculations OCS molecules were vertically excited
from the rovibronic-ground-state potential-energy surface (PES) to the $1\,^1\!A''$
($1\,^1\Sigma^{-}$) and the $2\,^1\!A'$ ($1\,^1\!\Delta$) states, \ie, the ground-state density was
projected onto these PESs. The PESs used are based on high-level MRCI/aug-cc-pVQZ electronic
structure calculations~\cite{Schmidt:JCP136:131101, Schmidt:JCP137:054313, McBane:JCP138:094313,
   Schmidt:ACP13:1511}; see ref.~\onlinecite{Schmidt:JCP136:131101, Schmidt:JCP137:054313} for
graphical representations. Following photoexcitation, the wavepackets were then propagated on these
PESs using the multiconfigurational time-dependent Hartree approach (MCTDH)~\cite{Meyer:CPL165:73,
   Manthe:JCP97:3199}. Jacobi coordinates as defined in \autoref{fig:coordinates} were used
throughout the quantum dynamics simulations, with $r$ denoting the CO bond length, $R$ the distance
of the S atom to the centre of mass of the carbon monoxide fragment, and $\theta$ the angle between
$r$ and $R$~\cite{Welsch:JCP136:064117, Welsch:MP110:703}. The wavepackets were propagated over
$400$~fs, using adaptive step sizes, with the photoexcitation taking place at $t=0$.

The \emph{ansatz} for a set of MCTDH~\cite{Meyer:CPL165:73} wave functions is given by
\onlinecite[(1) in ref.][]{Manthe:JCP97:3199}, with the so-called single-particle functions (SPF) as
time-dependent basis functions. The SPFs are expanded in a time-independed basis, see
\onlinecite[ref.][(2)]{Manthe:JCP97:3199}. The degrees of freedom in our calculations employed here
are the coordinates $R$, $r$, and $\theta$, as described above.

Employing fast-Fourier-transform (FFT) or discrete-variable-representation (DVR) schemes for
distances and angles, respectively, the SPFs were expanded in a time-independent basis. The
correlation DVR (CDVR)~\cite{Manthe:JCP105:6989} scheme was used to obtain matrix elements for
general potential energy surfaces.
\begin{table}[b]
   \centering%
   \begin{tabular}{ccccc}
     \hline\hline
     coordinate & expansion type & $n$ & $N$ & range \\
     \hline
     $R$ & FFT & 16 & 2048 & 2.8--72.0~a.u.\\
     $r$ &  FFT & 7 & 48 & 1.5--3.0 a.u. \\
     $\theta$ & Legendre-DVR & 22 & 128 & 0--$\pi$ \\
     \hline\hline
   \end{tabular}
   \caption{Expansion type, number of grid points $N$, and number of single-particle functions $n$
      used in the presented quantum dynamics simulations.}
   \label{tab:numbersreal}
\end{table}
Throughout this work, the revised constant mean-field scheme (CMF2)~\cite{Manthe:CP329:168} was used
to efficiently propagate the MCTDH wavepackets. The converged basis set sizes are given in
\autoref{tab:numbersreal}. Please note that slightly smaller grids are employed for calculating the
OCS ground state.

\begin{figure}
   \includegraphics[width=0.75\linewidth]{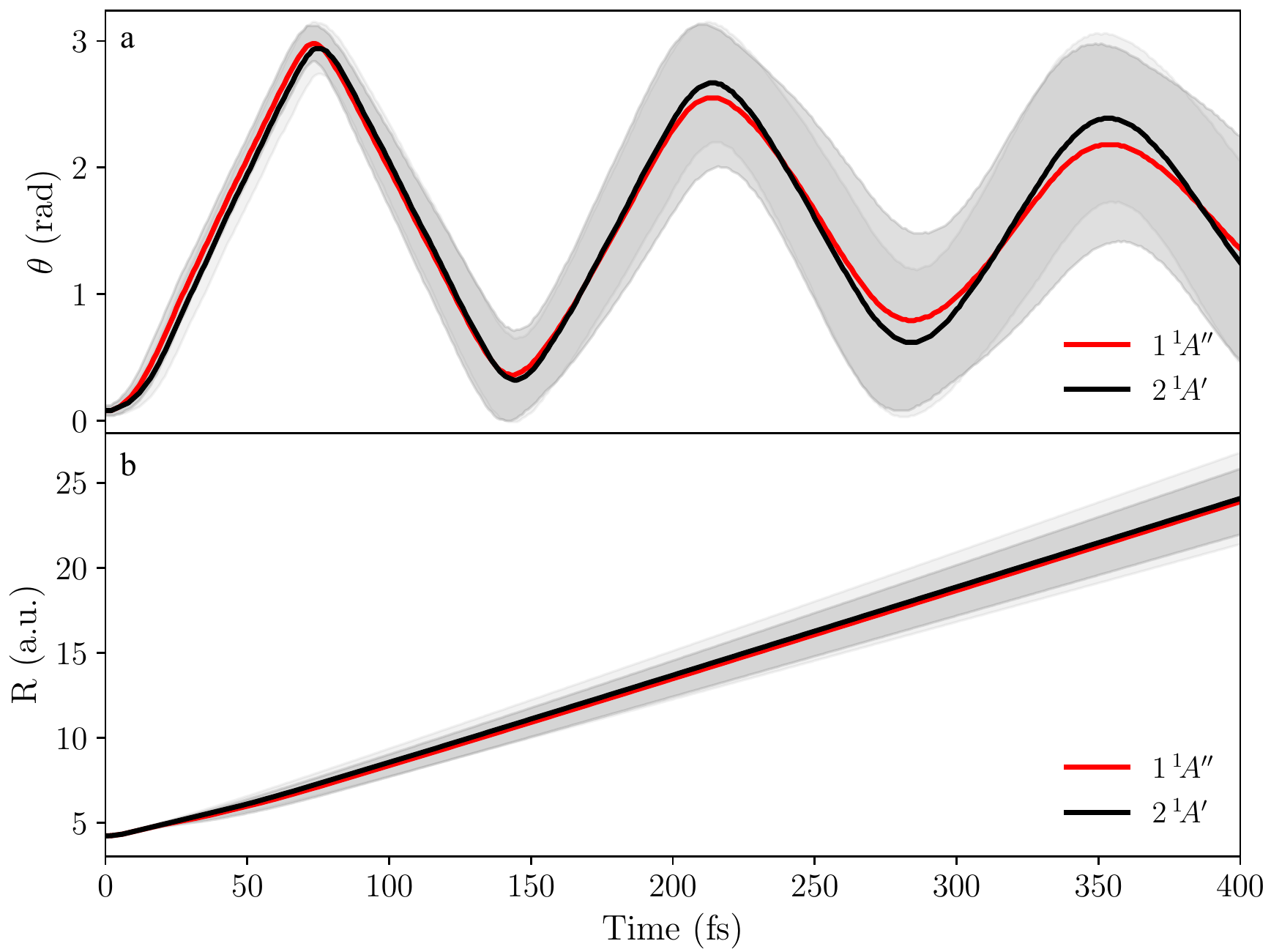}%
   \caption{Time-dependent expectation values of (a) $\theta$ and (b) $R$ for the computed
      wavepackets on the $1\,^1\!A''$ and $2\,^1\!A'$ potential-energy surfaces; see
      \autoref{fig:coordinates} and \autoref{sec:computation} for details. Standard deviations of
      the calculated probability densities are depicted as grey areas to show the spread of the
      wavepackets.}
   \label{fig:sim-expval}
\end{figure}
Movies of the wavepacket dynamics along the reaction coordinates obtained from our model in the
$2\,^1\!A'$ ($1\,^1\!\Delta$) and $1\,^1\!A''$ ($1\,^1\Sigma^{-}$) excited states are provided in
the supplementary material. \autoref{fig:sim-expval} shows the calculated time-dependent expectation
values of $\theta$ and $R$ for the calculated wavepacket dynamics on the $1\,^1\!A''$
($1\,^1\Sigma^{-}$) and $2\,^1\!A'$ ($1\,^1\!\Delta$) potential energy surfaces, which are very
similar. The temporal evolution of the wavepackets in these electronic states exhibits a monotonic
increase of $R$, \autoref[b]{fig:sim-expval}. For $t=100$~fs, $R$ is already twice the equilibrium
distance and from there on increases nearly linearly, implying that the chemical bond is effectively
being broken already at this time.

The time-dependent expectation values of $\theta$, see \autoref[a]{fig:sim-expval}, shows pronounced
oscillations with a period of $T=140$~fs. These oscillations can be assigned to the rotational
motion of the CO product resulting from the bending of the molecule following UV-excitation and
dissociation of OCS. The observed rotational period corresponds to the classical rotation period of
a state with rotational quantum number $J\approx61$, which fits previously
observed~\cite{Suzuki:JCP109:5778} terminal CO rotational-state distributions very well.

\section{Results and discussion}
\label{sec:results}
\begin{figure}[b]
   \includegraphics[width=\linewidth]{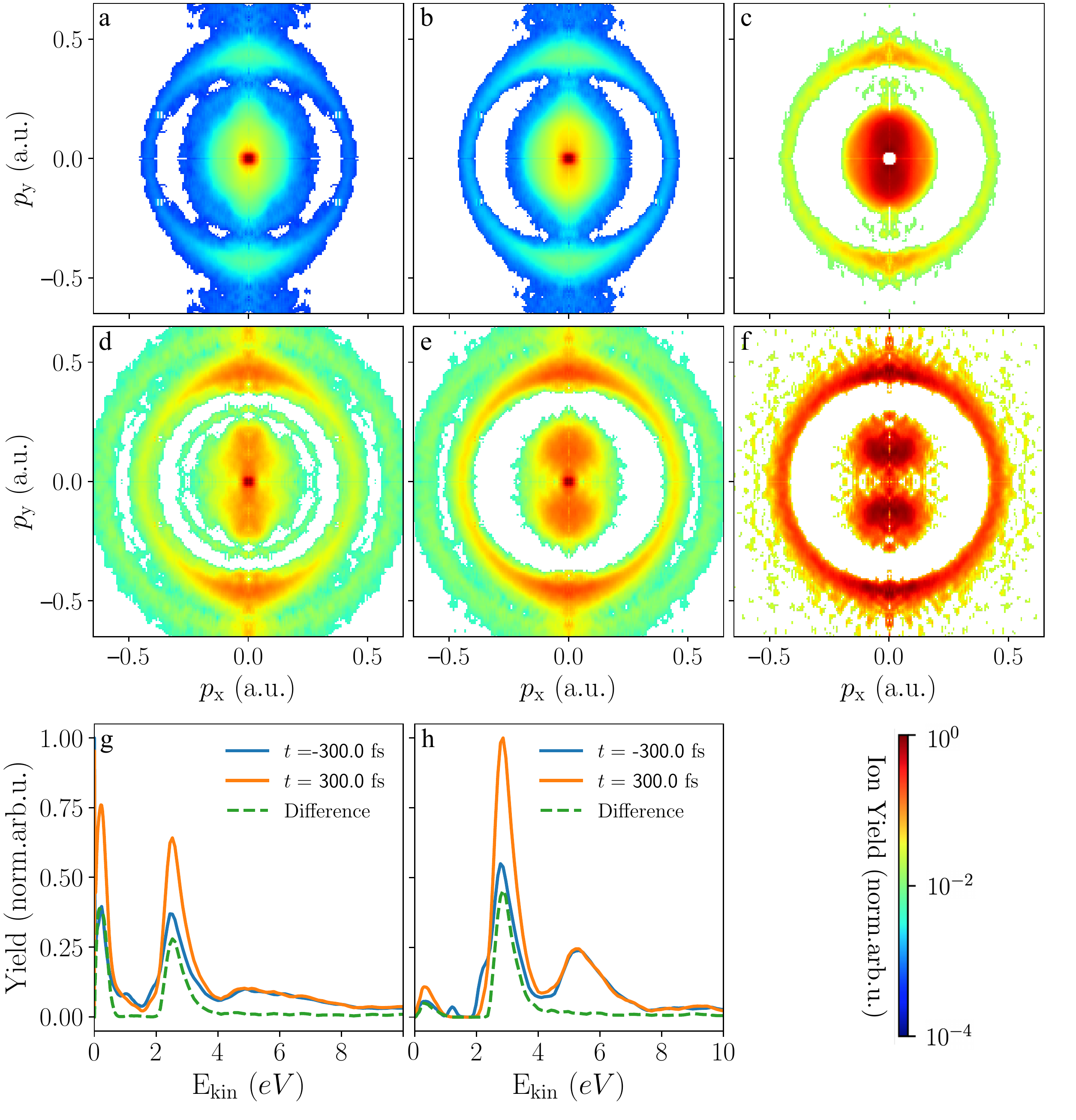}%
   \caption{Slices through 3D Abel-inverted ion-momentum distributions of S$^+$ and CO$^+$ fragments
      and corresponding angle-resolved kinetic energy spectra. (a--c,~g)~S$^+$ and (d--f,~h)~CO$^+$
      ion momentum distributions recorded at a UV-NIR time delay of (a,~d)~$\tau=-300$~fs and
      (b,~e)~$\tau=+300$~fs, (c,~f)~their differences, and (g,~h) the corresponding angle-integrated
      kinetic-energy distributions.}
   \label{fig:vmi_images}
\end{figure}

\autoref{fig:vmi_images} shows slices through the 3D ion-momentum distributions of
\hyperref[fig:vmi_images]{(a--c)} the S$^+$ and \hyperref[fig:vmi_images]{(d--f)} CO$^+$ fragments
recorded upon photoionisation of UV-excited OCS as well as \hyperref[fig:vmi_images]{(g,~h)} the
corresponding angle-integrated kinetic energy spectra.
In these measurements the pulse energies of the UV-excitation pulses (237~nm) and the NIR ionisation
pulses were set to 4.5~\uJ and 300~\uJ, respectively. The S$^+$ momentum distribution is composed of
two main channels that peak at kinetic energies of 0.22~eV and 2.53~eV. The 0.22~eV channel is
assigned to ionisation of OCS into low-energy excited states of the molecular OCS$^+$ cation, which
are known to dissociate to form a singly charged S atom and a neutral CO fragment in its electronic
ground state~\cite{Morse:IJMSIP184:67}. The 2.53~eV channel corresponds to Coulomb explosion
following double ionisation of the molecules by the NIR pulses leading to two singly charged S$^+$
and CO$^+$ fragments. In addition, a weaker and broader tail around 5~eV is observed, which is due
to Coulomb explosion with a doubly, or higher, charged CO$^{n+}(n\ge2)$ counterion.

Similarly, in the CO$^+$ angle-integrated kinetic energy spectrum, \autoref[h]{fig:vmi_images},
three channels are observed at kinetic energies of 0.33~eV, 2.84~eV, and 5.26~eV. Complementary to
S$^+$, the lowest energy channel can be attributed to the ionisation of the OCS molecule followed by
dissociation leading to a singly charged CO$^+$ fragment and a neutral S
atom~\cite{Morse:IJMSIP184:67}, whereas the two other channels can be assigned to Coulomb explosion
of the molecules by the NIR laser pulses with S$^+$ and S$^{2+}$ as counterions. Based on the
intensities of the lowest kinetic-energy peaks in the spectra in \autoref[g,~h]{fig:vmi_images}, the
dissociation of OCS$^+$ predominantly produces S$^+$ ions, in agreement with previous single photon
ionisation experiments~\cite{Morse:IJMSIP184:67}.

When the UV pulse precedes the NIR pulse a strong enhancement of the ion signal is observed for
both, S$^+$ and CO$^+$, fragments in the dissociative-ionisation channel and the first Coulomb
explosion channel, \ie, the channels with one or two overall charges in the system. This can be
rationalised by the fact that the UV-excited OCS has a much smaller ionisation energy (\Ei) and can
be ionised with significantly weaker fields/less NIR photons. The large increase of the signal
observed in the dissociative-ionisation channel of both fragments can \emph{a priori} be assigned to
UV-excitation to the $2\,^1\!A'$ ($1\,^1\!\Delta$) and $1\,^1\!A''$($1\,^1\Sigma^{-}$) states, which
dissociate quickly to form ground state CO$(X\,^1\Sigma^{+})$ and electronically excited
S$(^1\!D_{2})$ fragments~\cite{Sivakumar:JCP88:3692, Suzuki:JCP109:5778}, followed by ionisation of
one of the fragments by the NIR laser pulses. Considering that the ionisation energy of
S$(^1\!D_{2})$, $\Ei\approx9.2$~eV, is much lower than the one of CO$(X\,^1\Sigma^{+})$,
$\Ei\approx14$~eV, the NIR pulse is expected to preferentially ionise the S atom. This trend is
clearly observed experimentally as the dissociative-ionisation signal observed at positive time
delays is larger in the S$^+$ channel than in the CO$^+$ channel.

\begin{figure}
   \includegraphics[width=\linewidth]{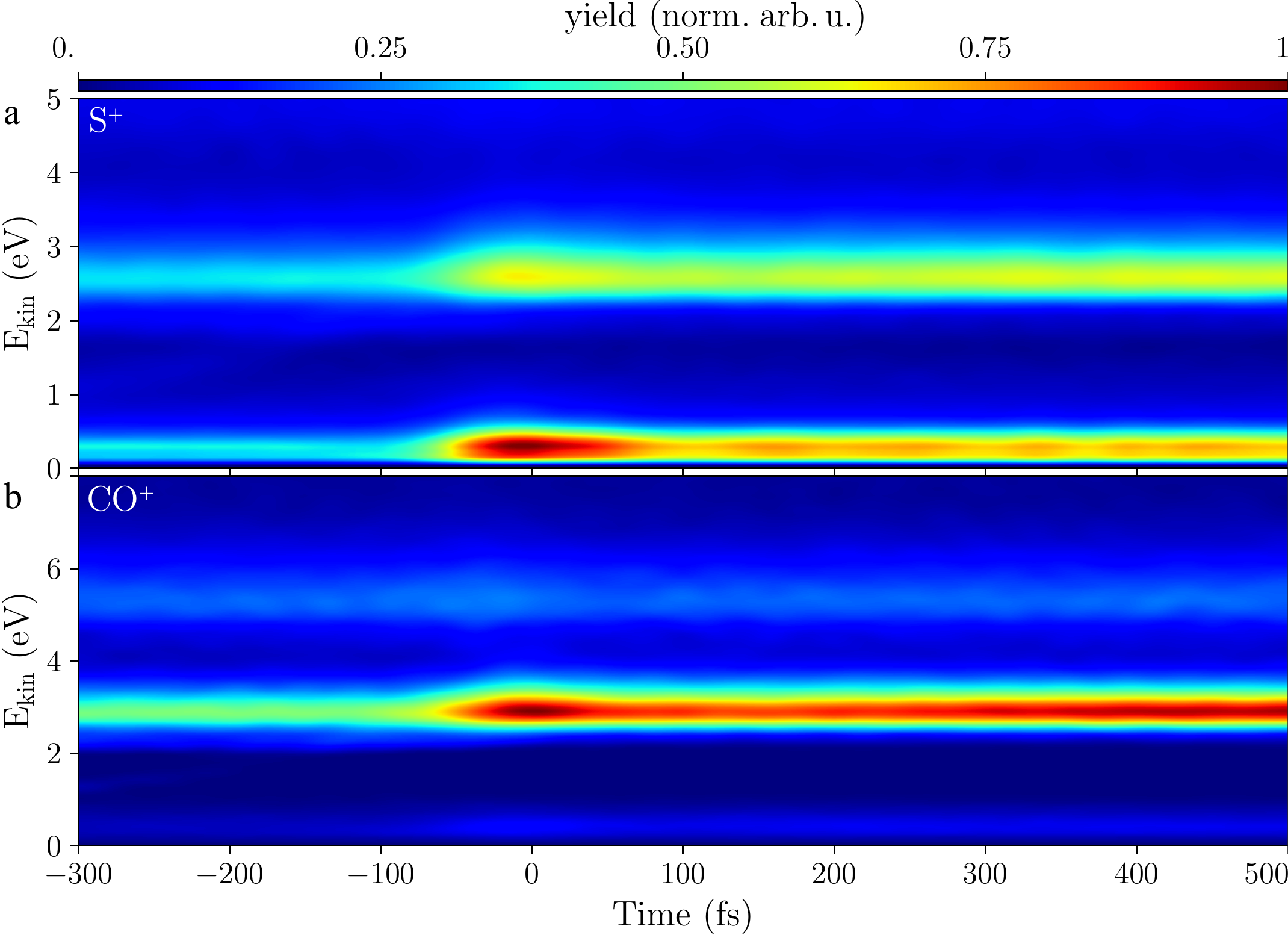}%
   \caption{Time-dependent angle-integrated kinetic energy spectra of (a) S$^+$ and (b) CO$^+$
      fragments in the time interval $\tau=-300\ldots500$~fs. A positive delay means the UV pulse
      precedes the NIR pulse.}
   \label{fig:td-KE-spectra}
\end{figure}

In principle, one might expect that the large increase of the signal in the first Coulomb explosion
channel similarly originates from the ionisation of two co-fragments formed following UV excitation
and dissociation of the neutral molecules. However, in the case of Coulomb explosion from
dissociating OCS molecules, we would expect the kinetic energy of the S$^+$ and CO$^+$ fragments to
decrease with increasing time delay, as the Coulomb repulsion energy decreases as $\ordsim1/R$ when
the internuclear distance $R$ increases~\cite{Stapelfeldt:PRA58:426}, which is not observed here,
see also \autoref{fig:td-KE-spectra}. Instead, the large increase of the ionisation yield in the
first Coulomb explosion channel suggests that a large number of OCS molecules remain bound after
UV-excitation.

The ion yield observed in the highest-kinetic energy channels, due to Coulomb explosion of an OCS
system with overall three (or more) charges, is identical at UV-NIR time delays of $\tau=-300$~fs
and $\tau=+300$~fs. This is seen in the differences of the momentum maps in
\autoref[c,~f]{fig:vmi_images}, where the corresponding outermost rings for these Coulomb explosion
channels are not present, and the corresponding kinetic energy spectra in
\autoref[g,~h]{fig:vmi_images}, where the spectra are identical at these high kinetic energies. This
could be due to the fact that these high-kinetic energy channels require the removal of at least
three electrons, including electrons from deeper orbitals than the HOMO, and, therefore, are little
affected by the initial UV-excitation of an electron from the HOMO orbital into unoccupied orbitals
of the neutral OCS.

In \autoref{fig:td-KE-spectra} the delay-dependent kinetic energy spectra for S$^+$ and CO$^+$
fragments recorded in the delay interval $\tau=-300\ldots500$~fs are shown. In the following
discussion we focus on the two lower-kinetic-energy channels of both ions, \ie, below 3~eV, as the
higher-energy channels only exhibit effects when the two pulses overlap in time.

A large enhancement of the ion signal is observed for both ionic fragments when the two pulses
overlap in time. This increase around $\tau=0$ is attributed to the increased efficiency of
two-color multiphoton ionisation when the two laser fields are present simultaneously. At positive
time delays, the signal levels for the two lowest ion kinetic energy channels observed in both
fragments remain nearly constant and are approximately two times larger than for negative delays.

The large increase of the dissociative-ionisation signal, \ie, in the lowest-kinetic-energy channels
at positive time delays, with the major contribution in S$^+$, provides information on the
dissociation dynamics in the UV-excited OCS molecule. As mentioned above, the main contribution is
from excitation to the $2\,^1\!A'$ ($1\,^1\!\Delta$) and $1\,^1\!A''$($1\,^1\Sigma^{-}$) states,
which quickly dissociate into S and CO. This was confirmed by the results of our wavepacket
calculations, which showed that at $t=100$~fs the bond is effectively broken, see
\autoref{sec:computation}. Thus, the NIR pulse will ionise the produced neutral S or CO fragments
with a preference for the S atom due to its much lower \Ei. We note that dissociation dynamics
occuring within the first 100~fs following UV-excitation cannot be unravelled in our measurements
due to the relatively long pulse duration ($\ordsim70$~fs) of the 237~nm pulse.

The time dependence of the Coulomb explosion channel resulting in S$^+$ and CO$^+$ with kinetic
energies of 2.53~eV and 2.84~eV, respectively, cannot be explained by NIR ionisation of neutral
fragments that result from prompt dissociation of the UV-excited OCS molecules: such a process would
have given rise to a kinetic energy spectra where the kinetic energy release decreases with
pump-probe time delay, which is not the case in the experimental data, see
\autoref{fig:td-KE-spectra}. Instead, we tentatively assign the large observed increase of the
Coulomb explosion channel for $\tau > 0$ to the UV-excitation of bound electronically excited
states, \eg, the $2\,^1\!A''$ ($1\,^1\!\Delta$), $2\,^3\!A'$ ($1\,^3\Delta$), and $2\,^3\!A''$
($1\,^3\Sigma^{-}$) triplet states, that are accessible at a photon energy of 237~nm. Due to their
lower \Ei, these excited OCS molecules are readily multiply ionised by the probe IR pulse, enhancing
their contribution in the kinetic-energy spectra. A similarly increased probability for dissociative
single ionisation could also partly be responsible for the increased S$^+$ and CO$^+$ signals
observed in the lowest kinetic-energy channels.

\begin{figure}
   \includegraphics[width=0.85\linewidth]{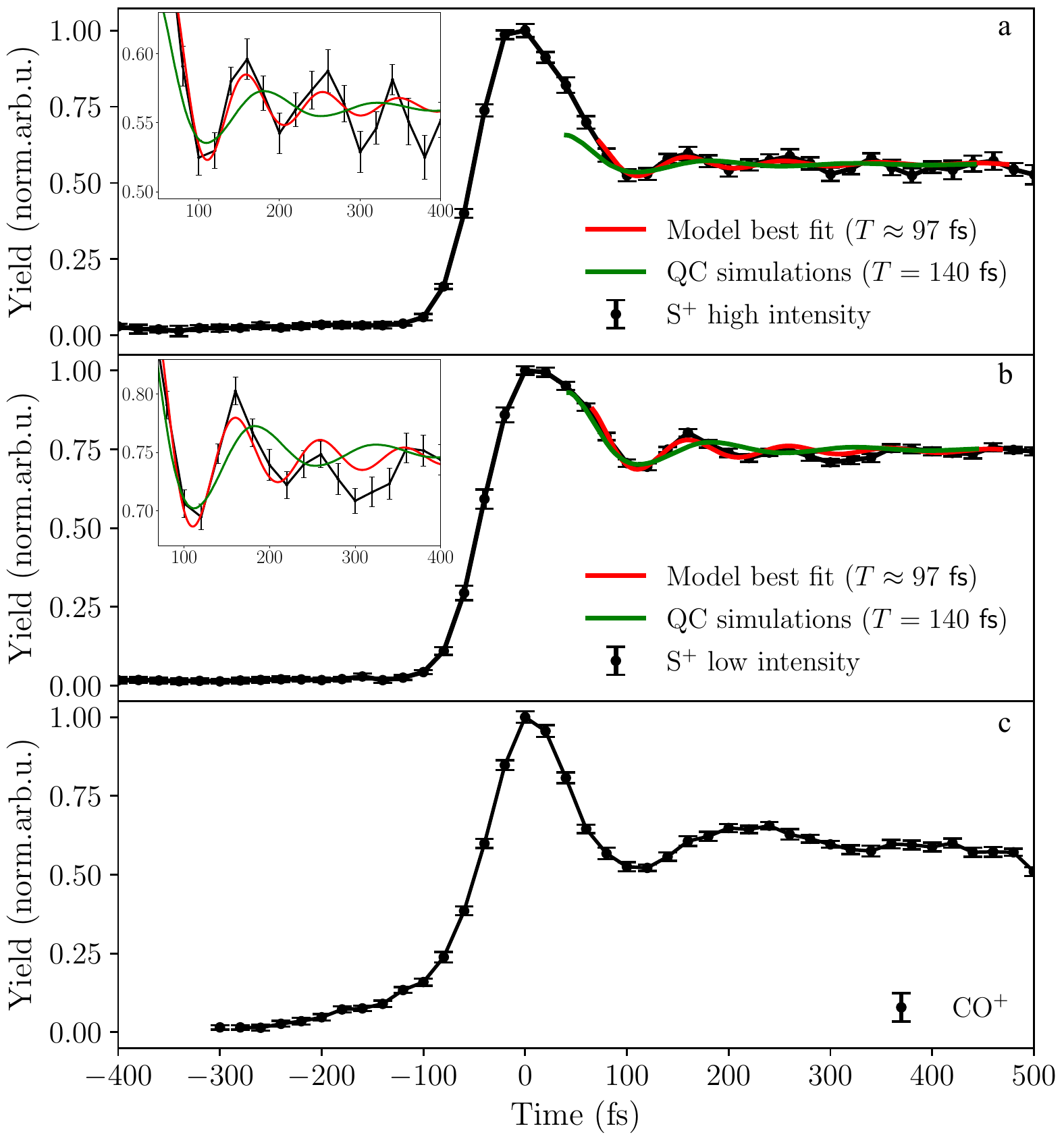}%
   \caption{Time-dependent S$^+$ and CO$^+$ yields for the delay range $\tau=-400\ldots500$~fs.
      (a,~b)~S$^+$~ions with a kinetic energy of $\ordsim0.22$~eV for UV and NIR pulse energies of
      (a) 10~\uJ and 1.3~\mJ and (b)~4.5~\uJ and 300~\uJ, respectively. The (black) experimental
      S$^{+}$ yields are compared to (red and green) simulations considering a modified ADK
      ionisation rate for the sulfur atom due to the rotational motion of the CO fragment, see the
      text for details. The insets show a magnified view on the oscillations for better visibility.
      (c)~CO$^+$-ions with a kinetic energy of $\ordsim0.38$~eV for UV and NIR pulse energies of
      10~\uJ and 1.3~\mJ, respectively.}
   \label{fig:oscillations}
\end{figure}
In addition to the large increase of the S$^+$ and CO$^+$ ionisation yields, weak but pronounced
oscillations are observed at positive time delays in the dissociative-ionisation channels of both
S$^+$ and CO$^+$. Integrating the ionisation yields around the lowest kinetic energy channels
observed in the S$^+$ and CO$^{+}$ fragments, \ie, at kinetic energies of $0.22$~eV and $0.33$~eV
yields the time-dependent ion yields in \autoref{fig:oscillations}. The oscillations exhibit a
period of $T_{\text{S}^{+}}\approx100$~fs for S$^+$. For the CO$^+$ channel the oscillation period
seems to be longer, but no well-defined value could be extracted as no full period is observed
within the time interval that the CO$^+$ fragments were recorded. The observed oscillations are
independent of the UV and NIR pulse energies used, which suggests that the observed dynamics is not
due to two-photon excitations to higher excited states of neutral OCS.

The dissociation dynamics of OCS in the $1\,^1\!A''$ and $2\,^1\!A'$ potential energy surfaces is
known to involve large bending motion of the molecule, which results in CO products with high
rotational excitation \cite{Sivakumar:JCP88:3692, Suzuki:JCP109:5778}. This rotational excitation
was also clearly observed in our wavepacket calculations, \ie, the time-dependent expectation value
of $\theta$ shown in \autoref[a]{fig:sim-expval}. Building upon these computational results we
attempt to explain the oscillations observed in ion yields, esp.\ the time-dependent S$^+$ yields
using an intuitive physical picture: One would expect that during dissociation the dipole moment of
the rotating CO influences the instantaneous ionisation rate of the nearby S atom: the direction of
the rotating dipole with respect to the $R$ coordinate shifts the ionisation potential of the sulfur
atom, and thus its ionisation rate. To assess whether this effect is responsible for the observed
oscillations, we used a simple ADK-tunnel-ionisation model~\cite{Ammosov:SVJETP64:1191} to simulate
the expected modulation of the S-atom ionisation rate as a function of time under the influence of
the rotating CO dipole. For time delays $\tau > 100$~fs, approximately the time-scale for the onset
of the observed oscillations, $R\gtrsim9$~\au and thus S can be considered to be a quasi-free atom.
The ionisation signal can then be well approximated by the strong-field-ionisation rate of S in the
field of the CO dipole moment. We calculated the electric potential of the CO dipole moment at the
position of the S atom as $V(t)=\mu\cos\theta(t)/R(t)$ with the angle $\theta(t)$ of the rotating CO
fragment at time $t$ and the dipole moment of CO $\mu=0.122$~D. We note that this expression only
holds for $R\gg{r}$, see \autoref{fig:coordinates} for the coordinates. However, we also benchmarked
the results against further multipole expansions and even explicit partial charges on C and O, which
did not make any difference. For $R\approx10$~\au the change of \Ei is on the order of 0.1~eV. The
time-dependent ionisation rate was computed once using the time-dependent expectation values for
$\theta(t)$ from our wavepacket simulations, \autoref[a]{fig:sim-expval}, and once by fitting the
time-period for the rotational motion of CO to the experimental curves by employing a least-squares
minimisation. In both cases, the time-dependent change of the CO--S internuclear distance $R(t)$ was
taken directly from the results of our wavepacket simulations on the $1\,^1\!A''$ and $2\,^1\!A'$
potential energy surfaces, see \autoref[b]{fig:sim-expval}.

The results of this model are shown in \autoref[a,~b]{fig:oscillations}, in comparison with the
experimentally measured time-dependent S$^+$ dissociative ionisation yields recorded with (a) low
and (b) high UV and NIR pulse energies, respectively. The rotational period of $T=140$~fs, obtained
from the quantum chemistry simulations, directly translates into a $T_{\text{S}^+}=140$~fs period in
the ionisation signal and, therefore, does not match the measured oscillations. When used as a
fitting parameter, a rotational period of $T=97_{-8}^{+14}$~fs is retrieved for the two
intensities. This fast rotational period would correspond to the population of rotationally excited
CO fragments with $J>80$, which clearly disagrees with earlier REMPI spectra of the S
photoproducts~\cite{Sivakumar:JPC89:3609, Sato:JPC99:16307, Suzuki:JCP109:5778,
   Katayanagi:CPL247:571, Lipciuc:PCCP13:8549, Sofikitis:PRL118:253001, Sofikitis:PRA98:033417}.
Therefore, we conclude that the observed oscillations are not caused by the electrostatic
interaction of the highly rotationally excited CO fragment with the neutral S atom along the
dissociation pathway.

Instead, the fast oscillations observed in the S$^+$ products of the UV-induced dissociation can be
explained by the population of vibrational wavepackets in one or more electronically excited states
of OCS, for which bound vibrational states are accessible at 237~nm. REMPI spectra of the S
photoproducts showed a broad, unstructured spectrum in the major S$(^1\!D_{2})$
channel~\cite{Toulson:JPCA120:6745}. This suggests that the absorption spectrum and correspondingly
the observed products are dominated by excitation to the repulsive $2\,^1\!A'$ ($1\,^1\!\Delta$) and
$1\,^1\!A''$ ($1\,^1\Sigma^{-}$) states, which lead to prompt dissociation, see also
\autoref{fig:sim-expval}. The repulsive nature of these two states along the dissociation coordinate
was confirmed by earlier calculations~\cite{Schmidt:JCP137:054313} However, these calculations also
indicated that these states have shallow local minima supporting bound vibrational states around
bending angles of $\theta\approx\degree{40}$. The profound vibrational structure observed in the
low-energy tail of the absorption spectrum of OCS, \eg, above $\ordsim270$~nm, was assigned to these
states~\cite{Toulson:JPCA120:6745}. Although in our experiment we excite the OCS molecules to much
higher energies, we cannot fully rule out the formation of vibrational wavepackets in these
$2\,^1\!A'$ ($1\,^1\!\Delta$) and $1\,^1\!A''$ ($1\,^1\Sigma^{-}$) states, which would predissociate
through non-adiabatic coupling to the $1\,^1\!A'$ ground state.

In addition, a strong vibrational progression superimposed on a broad continuum was recently
observed in the REMPI spectrum of the minor S$(^3\!P_{J})$ products following UV-excitation of OCS
in its first absorption band~\cite{Toulson:JPCA120:6745}. The diffuse vibrational structure was
assigned to direct excitation to the bound $2\,^1\!A''$ ($1\,^1\!\Delta$) and $2\,^3\!A''$
($1\,^3\Sigma^{-}$) metastable states. The $2\,^3\!A''$ ($1\,^3\Sigma^{-}$) triplet state can
predissociate through spin-orbit coupling to the dissociative $2\,^1\!A'$ ($1\,^1\!\Delta$) state,
with subsequent intersystem crossing to form S$(^3\!P_{J})$ products. Similarly, the $2\,^1\!A''$
($1\,^1\Delta$) state can predissociate through rovibronic coupling to the dissociative $1\,^1\!A''$
($1\,^1\Sigma^{-}$), with subsequent intersystem crossing to the $2\,^3\!A''$ ($1\,^3\Sigma^-$) to
form S$(^3\!P_{J})$ products at larger $R$. Interestingly, a weak absorption feature was observed at
a UV wavelength of 236.12~nm, \ie, in the central part of our UV pump-laser wavelength, in the REMPI
spectrum of the minor S$(^3\!P_{J})$ product~\cite{Toulson:JPCA120:6745}. This band was assigned to
the C--S stretching vibration $\nu_1$ of the $2\,^1\!A''$ ($1\,^1\!\Delta$) state. It strongly
suggests that the fast oscillations of the S$^+$ and CO$^+$ products in our time-resolved
measurements are due to vibrational wavepacket dynamics in the $2\,^1\!A''$ ($1\,^1\!\Delta$)
states, which then predissociate to form S$(^3\!P_{J})$ products. This UV excitation to bound states
is also corroborated by the UV-induced enhancements in the Coulomb-explosion channels discussed
above.

Further experiments and theoretical investigations are clearly needed to draw definite conclusions
on the underlying physical process leading to the observed oscillations in the S$^+$ signal. We are
currently experimentally investigating the OCS reaction system using laser-induced electron
diffraction in the molecular frame~\cite{Karamatskos:NatComm10:3364, Karamatskos:JCP150:244301,
   Trabattoni:NatComm11:2546, Karamatskos:thesis:2019}. Imaging the dynamics observed here with our
previously achieved 5~pm spatial~\cite{Karamatskos:JCP150:244301} and sub-100~fs temporal resolution
would provide a very direct structural view on the ongoing dynamics. Alternatively, MeV ultrafast
electron diffraction with sufficient time resolution is also within reach~\cite{Wolf:NatChem11:504}
and could provide significant information on this structurally simple model system.

Complementary theoretical studies would undoubtedly be needed as well. In the vicinity of the
Franck-Condon point, at least eight electronic states are known to be involved in the dissociation
dynamics of OCS (four singlets and four triplets)~\cite{Schmidt:JCP136:131101}. Additional
electronic states correlate to the two relevant dissociation limits, S$(^1\!D_{2})$ and
S$(^3\!P_{J})$, which may become involved as the system dissociates. Due to the presence of the
second-main-row S atom, the spin-orbit and non-adiabatic couplings are expected to be comparable,
leading to both, direct excitation of the triplet manifold~\cite{Schmidt:JCP136:131101} and rapid
inter-system crossing and re-crossing. In the related, simpler case of bound dynamics in SO$_2$, the
non-adiabatic and spin-orbit coupling were found to be intertwined~\cite{Wilkinson:JCP140:204301}
and both had to be treated to reach an agreement with experiment~\cite{Mai:JCP140:204302}.
Furthermore, the energy surfaces, and especially the transition dipoles in OCS are known to be
highly sensitive to the treatment of electron correlation~\cite{McBane:JCP138:094313}, potentially
requiring correlation treatment at the coupled-cluster-triples level to achieve a satisfactory
agreement with the experiment~\cite{Schmidt:JCP141:184310}. Finally, at the high rotational
excitation levels observed for the CO dissociation products, the reduced-dimensionality model may
become insufficient~\cite{Owens:PRL121:193201}. Although challenging, the multi-surface,
full-dimensional description of the dissociation dynamics of photoexcited OCS is within reach of the
\emph{ab initio} electronic structure and MCTDH simulations.

\section{Conclusions and outlook}
\label{sec:conclusion}
The ultrafast photodissociation dynamics of OCS following UV-photoexcitation at
\mbox{$\lambda=237$~nm} was investigated by time-resolved ion-momentum spectroscopy using a strong,
time-delayed, 790~nm laser pulse. We observed a strong enhancement of the ion yields in the S$^+$
and CO$^+$ fragments due to UV excitation. This is consistent with the excitation of the molecules
to the repulsive $2\,^1\!A'$ ($1\,^1\!\Delta$) and $1\,^1\!A''$ ($1\,^1\Sigma^{-}$) states followed
by prompt dissociation leading to S$(^1\!D)$ and rotationally excited CO$(X\,^1\Sigma^{+})$
photofragments. A pronounced increase of the Coulomb explosion ionisation rate following UV
excitation of the molecules was also observed. This was assigned to the excitation of molecules in
bound electronically excited states, most likely the $2\,^1\!A''$ ($1\,^1\!\Delta$) and $2\,^3\!A''$
($1\,^3\Sigma^{-}$) states. From a time delay of 100 fs onward fast oscillations of the S$^+$ and
CO$^+$ yields were observed on top of a large time-independent signal, with a well-defined period of
100~fs in the S$^+$ signals.

Furthermore, we performed wavepacket dynamics simulations on the $2\,^1\!A'$ ($1\,^1\!\Delta$) and
$1\,^1\!A''$ ($1\,^1\Sigma^{-}$) states, which showed direct dissociation with fast rotation of the
CO products with a 140~fs period. Using a simple electrostatic interaction and
strong-field-ionisation model, we could rule out that this CO rotation produces the observed
oscillations. Instead, these oscillations might reflect vibrational wavepacket dynamics in the
electronically excited bound $2\,^1\!A''$ ($1\,^1\!\Delta$) and $2\,^3\!A''$ ($1\,^3\Sigma^{-}$)
metastable states, which can predissociate by spin-orbit or rovibronic couplings to the minor
dissociation channel leading to ground state S$(^3\!P_{J})$. This is in agreement with pronounced
vibrational progressions in REMPI spectra of the S$(^3\!P_{J})$ products in the same
excitation-energy range~\cite{Toulson:JPCA120:6745}. Clearly, additional wavepacket dynamics
simulations including the different electronic states that are known to be involved in the
dissociation dynamics of OCS, as well as their couplings, are needed to understand the mechanism
responsible for the observed fast oscillations of the S$^+$ dissociative ion yield.

This first time-resolved investigation of the photodissosiation dynamics of OCS clearly demonstrates
the need for further advanced experimental and theoretical studies to elucidate the underlying
physical processes responsible for the fast oscillations observed experimentally. Time-resolved
electron-ion coincidence spectroscopy with extreme-ultraviolet laser pulses as a
probe~\cite{Pathak:NatChem12:795} could provide further information on the investigated ultrafast
dissociation process. Furthermore, experimental studies providing atomically-resolved images of the
structural changes during these UV-induced dynamics could be very valuable. As a first step toward
this goal, we have recently retrieved the structure of OCS in its equilibrium geometry using
laser-induced electron diffraction~\cite{Karamatskos:JCP150:244301} and time-resolved LIED studies
of the OCS photodissociation dynamics are underway to provide a first molecular movie of this
benchmark system.

\section*{Acknowledgements}
We thank Johan Schmidt for providing us with the numerical data of the potential energy surfaces
used in our wavepacket calculations.

This work has been supported by the Deutsche Forschungsgemeinschaft (DFG) through the priority
program ``Quantum Dynamics in Tailored Intense Fields'' (QUTIF, SPP1840, AR 4577/4, KU 1527/3) and
by the Cluster of Excellence ``Advanced Imaging of Matter'' (AIM, EXC 2056, ID~390715994) of the
Deutsche Forschungsgemeinschaft.

\section*{Author contributions}
J.K.\ and A.R.\ conceived and planned the experiment, E.T.K., S.Y., and A.R.\ performed the experiment, E.T.K.\
and S.Y.\ analysed the data, R.W.\ performed the wavepacket calculations, E.T.K., S.P., R.W., J.K.,
and A.R.\ interpreted the findings, E.T.K., J.K., and A.R.\ drafted the manuscript, and all authors
discussed the results and contributed to the final manuscript.

\section*{Conflicts of interest}
The authors declare no conflicts of interest.

\bibliography{string,cmi}

\begin{thebibliography}{77}%
\makeatletter
\providecommand \@ifxundefined [1]{%
 \@ifx{#1\undefined}
}%
\providecommand \@ifnum [1]{%
 \ifnum #1\expandafter \@firstoftwo
 \else \expandafter \@secondoftwo
 \fi
}%
\providecommand \@ifx [1]{%
 \ifx #1\expandafter \@firstoftwo
 \else \expandafter \@secondoftwo
 \fi
}%
\providecommand \natexlab [1]{#1}%
\providecommand \enquote  [1]{``#1''}%
\providecommand \bibnamefont  [1]{#1}%
\providecommand \bibfnamefont [1]{#1}%
\providecommand \citenamefont [1]{#1}%
\providecommand \href@noop [0]{\@secondoftwo}%
\providecommand \href [0]{\begingroup \@sanitize@url \@href}%
\providecommand \@href[1]{\@@startlink{#1}\@@href}%
\providecommand \@@href[1]{\endgroup#1\@@endlink}%
\providecommand \@sanitize@url [0]{\catcode `\\12\catcode `\$12\catcode
  `\&12\catcode `\#12\catcode `\^12\catcode `\_12\catcode `\%12\relax}%
\providecommand \@@startlink[1]{}%
\providecommand \@@endlink[0]{}%
\providecommand \url  [0]{\begingroup\@sanitize@url \@url }%
\providecommand \@url [1]{\endgroup\@href {#1}{\urlprefix }}%
\providecommand \urlprefix  [0]{URL }%
\providecommand \Eprint [0]{\href }%
\providecommand \doibase [0]{https://doi.org/}%
\providecommand \selectlanguage [0]{\@gobble}%
\providecommand \bibinfo  [0]{\@secondoftwo}%
\providecommand \bibfield  [0]{\@secondoftwo}%
\providecommand \translation [1]{[#1]}%
\providecommand \BibitemOpen [0]{}%
\providecommand \bibitemStop [0]{}%
\providecommand \bibitemNoStop [0]{.\EOS\space}%
\providecommand \EOS [0]{\spacefactor3000\relax}%
\providecommand \BibitemShut  [1]{\csname bibitem#1\endcsname}%
\let\auto@bib@innerbib\@empty
\bibitem [{\citenamefont {Zewail}(2000)}]{Zewail:JPCA104:5660}%
  \BibitemOpen
  \bibfield  {author} {\bibinfo {author} {\bibfnamefont {A.~H.}\ \bibnamefont
  {Zewail}},\ }\bibfield  {title} {\enquote {\bibinfo {title}
  {{F}emtochemistry: {A}tomic-{S}cale {D}ynamics of the {C}hemical {B}ond},}\
  }\href {https://doi.org/10.1021/jp001460h} {\bibfield  {journal} {\bibinfo
  {journal} {J. Phys. Chem. A}\ }\textbf {\bibinfo {volume} {104}},\ \bibinfo
  {pages} {5660--5694} (\bibinfo {year} {2000})}\BibitemShut {NoStop}%
\bibitem [{\citenamefont {Young}\ \emph {et~al.}(2018)\citenamefont {Young},
  \citenamefont {Ueda}, \citenamefont {Gühr}, \citenamefont {Bucksbaum},
  \citenamefont {Simon}, \citenamefont {Mukamel}, \citenamefont {Rohringer},
  \citenamefont {Prince}, \citenamefont {Masciovecchio}, \citenamefont {Meyer},
  \citenamefont {Rudenko}, \citenamefont {Rolles}, \citenamefont {Bostedt},
  \citenamefont {Fuchs}, \citenamefont {Reis}, \citenamefont {Santra},
  \citenamefont {Kapteyn}, \citenamefont {Murnane}, \citenamefont {Ibrahim},
  \citenamefont {Légaré}, \citenamefont {Vrakking}, \citenamefont {Isinger},
  \citenamefont {Kroon}, \citenamefont {Gisselbrecht}, \citenamefont
  {L’Huillier}, \citenamefont {Wörner}, and\ \citenamefont
  {Leone}}]{Young:JPB51:032003}%
  \BibitemOpen
  \bibfield  {author} {\bibinfo {author} {\bibfnamefont {L.}~\bibnamefont
  {Young}}, \bibinfo {author} {\bibfnamefont {K.}~\bibnamefont {Ueda}},
  \bibinfo {author} {\bibfnamefont {M.}~\bibnamefont {Gühr}}, \bibinfo
  {author} {\bibfnamefont {P.~H.}\ \bibnamefont {Bucksbaum}}, \bibinfo {author}
  {\bibfnamefont {M.}~\bibnamefont {Simon}}, \bibinfo {author} {\bibfnamefont
  {S.}~\bibnamefont {Mukamel}}, \bibinfo {author} {\bibfnamefont
  {N.}~\bibnamefont {Rohringer}}, \bibinfo {author} {\bibfnamefont {K.~C.}\
  \bibnamefont {Prince}}, \bibinfo {author} {\bibfnamefont {C.}~\bibnamefont
  {Masciovecchio}}, \bibinfo {author} {\bibfnamefont {M.}~\bibnamefont
  {Meyer}}, \bibinfo {author} {\bibfnamefont {A.}~\bibnamefont {Rudenko}},
  \bibinfo {author} {\bibfnamefont {D.}~\bibnamefont {Rolles}}, \bibinfo
  {author} {\bibfnamefont {C.}~\bibnamefont {Bostedt}}, \bibinfo {author}
  {\bibfnamefont {M.}~\bibnamefont {Fuchs}}, \bibinfo {author} {\bibfnamefont
  {D.~A.}\ \bibnamefont {Reis}}, \bibinfo {author} {\bibfnamefont
  {R.}~\bibnamefont {Santra}}, \bibinfo {author} {\bibfnamefont
  {H.}~\bibnamefont {Kapteyn}}, \bibinfo {author} {\bibfnamefont
  {M.}~\bibnamefont {Murnane}}, \bibinfo {author} {\bibfnamefont
  {H.}~\bibnamefont {Ibrahim}}, \bibinfo {author} {\bibfnamefont
  {F.}~\bibnamefont {Légaré}}, \bibinfo {author} {\bibfnamefont {M.~J.~J.}\
  \bibnamefont {Vrakking}}, \bibinfo {author} {\bibfnamefont {M.}~\bibnamefont
  {Isinger}}, \bibinfo {author} {\bibfnamefont {D.}~\bibnamefont {Kroon}},
  \bibinfo {author} {\bibfnamefont {M.}~\bibnamefont {Gisselbrecht}}, \bibinfo
  {author} {\bibfnamefont {A.}~\bibnamefont {L’Huillier}}, \bibinfo {author}
  {\bibfnamefont {H.~J.}\ \bibnamefont {Wörner}}, and\ \bibinfo {author}
  {\bibfnamefont {S.~R.}\ \bibnamefont {Leone}},\ }\bibfield  {title} {\enquote
  {\bibinfo {title} {Roadmap of ultrafast x-ray atomic and molecular
  physics},}\ }\href {https://doi.org/10.1088/1361-6455/aa9735} {\bibfield
  {journal} {\bibinfo  {journal} {J. Phys. B}\ }\textbf {\bibinfo {volume}
  {51}},\ \bibinfo {pages} {032003} (\bibinfo {year} {2018})}\BibitemShut
  {NoStop}%
\bibitem [{\citenamefont {Barty}, \citenamefont {K{\"u}pper}, and\
  \citenamefont {Chapman}(2013)}]{Barty:ARPC64:415}%
  \BibitemOpen
  \bibfield  {author} {\bibinfo {author} {\bibfnamefont {A.}~\bibnamefont
  {Barty}}, \bibinfo {author} {\bibfnamefont {J.}~\bibnamefont {K{\"u}pper}},\
  and\ \bibinfo {author} {\bibfnamefont {H.~N.}\ \bibnamefont {Chapman}},\
  }\bibfield  {title} {\enquote {\bibinfo {title} {Molecular imaging using
  x-ray free-electron lasers},}\ }\href
  {https://doi.org/10.1146/annurev-physchem-032511-143708} {\bibfield
  {journal} {\bibinfo  {journal} {Annu. Rev. Phys. Chem.}\ }\textbf {\bibinfo
  {volume} {64}},\ \bibinfo {pages} {415--435} (\bibinfo {year}
  {2013})}\BibitemShut {NoStop}%
\bibitem [{\citenamefont {Pande}\ \emph {et~al.}(2016)\citenamefont {Pande},
  \citenamefont {Hutchison}, \citenamefont {Groenhof}, \citenamefont {Aquila},
  \citenamefont {Robinson}, \citenamefont {Tenboer}, \citenamefont {Basu},
  \citenamefont {Boutet}, \citenamefont {DePonte}, \citenamefont {Liang},
  \citenamefont {White}, \citenamefont {Zatsepin}, \citenamefont {Yefanov},
  \citenamefont {Morozov}, \citenamefont {Oberthuer}, \citenamefont {Gati},
  \citenamefont {Subramanian}, \citenamefont {James}, \citenamefont {Zhao},
  \citenamefont {Koralek}, \citenamefont {Brayshaw}, \citenamefont {Kupitz},
  \citenamefont {Conrad}, \citenamefont {Roy-Chowdhury}, \citenamefont {Coe},
  \citenamefont {Metz}, \citenamefont {Xavier}, \citenamefont {Grant},
  \citenamefont {Koglin}, \citenamefont {Ketawala}, \citenamefont {Fromme},
  \citenamefont {{\v S}rajer}, \citenamefont {Henning}, \citenamefont {Spence},
  \citenamefont {Ourmazd}, \citenamefont {Schwander}, \citenamefont
  {Weierstall}, \citenamefont {Frank}, \citenamefont {Fromme}, \citenamefont
  {Barty}, \citenamefont {Chapman}, \citenamefont {Moffat}, \citenamefont {van
  Thor}, and\ \citenamefont {Schmidt}}]{Pande:Science352:725}%
  \BibitemOpen
  \bibfield  {author} {\bibinfo {author} {\bibfnamefont {K.}~\bibnamefont
  {Pande}}, \bibinfo {author} {\bibfnamefont {C.~D.~M.}\ \bibnamefont
  {Hutchison}}, \bibinfo {author} {\bibfnamefont {G.}~\bibnamefont {Groenhof}},
  \bibinfo {author} {\bibfnamefont {A.}~\bibnamefont {Aquila}}, \bibinfo
  {author} {\bibfnamefont {J.~S.}\ \bibnamefont {Robinson}}, \bibinfo {author}
  {\bibfnamefont {J.}~\bibnamefont {Tenboer}}, \bibinfo {author} {\bibfnamefont
  {S.}~\bibnamefont {Basu}}, \bibinfo {author} {\bibfnamefont {S.}~\bibnamefont
  {Boutet}}, \bibinfo {author} {\bibfnamefont {D.~P.}\ \bibnamefont {DePonte}},
  \bibinfo {author} {\bibfnamefont {M.}~\bibnamefont {Liang}}, \bibinfo
  {author} {\bibfnamefont {T.~A.}\ \bibnamefont {White}}, \bibinfo {author}
  {\bibfnamefont {N.~A.}\ \bibnamefont {Zatsepin}}, \bibinfo {author}
  {\bibfnamefont {O.}~\bibnamefont {Yefanov}}, \bibinfo {author} {\bibfnamefont
  {D.}~\bibnamefont {Morozov}}, \bibinfo {author} {\bibfnamefont
  {D.}~\bibnamefont {Oberthuer}}, \bibinfo {author} {\bibfnamefont
  {C.}~\bibnamefont {Gati}}, \bibinfo {author} {\bibfnamefont {G.}~\bibnamefont
  {Subramanian}}, \bibinfo {author} {\bibfnamefont {D.}~\bibnamefont {James}},
  \bibinfo {author} {\bibfnamefont {Y.}~\bibnamefont {Zhao}}, \bibinfo {author}
  {\bibfnamefont {J.}~\bibnamefont {Koralek}}, \bibinfo {author} {\bibfnamefont
  {J.}~\bibnamefont {Brayshaw}}, \bibinfo {author} {\bibfnamefont
  {C.}~\bibnamefont {Kupitz}}, \bibinfo {author} {\bibfnamefont
  {C.}~\bibnamefont {Conrad}}, \bibinfo {author} {\bibfnamefont
  {S.}~\bibnamefont {Roy-Chowdhury}}, \bibinfo {author} {\bibfnamefont {J.~D.}\
  \bibnamefont {Coe}}, \bibinfo {author} {\bibfnamefont {M.}~\bibnamefont
  {Metz}}, \bibinfo {author} {\bibfnamefont {P.~L.}\ \bibnamefont {Xavier}},
  \bibinfo {author} {\bibfnamefont {T.~D.}\ \bibnamefont {Grant}}, \bibinfo
  {author} {\bibfnamefont {J.~E.}\ \bibnamefont {Koglin}}, \bibinfo {author}
  {\bibfnamefont {G.}~\bibnamefont {Ketawala}}, \bibinfo {author}
  {\bibfnamefont {R.}~\bibnamefont {Fromme}}, \bibinfo {author} {\bibfnamefont
  {V.}~\bibnamefont {{\v S}rajer}}, \bibinfo {author} {\bibfnamefont
  {R.}~\bibnamefont {Henning}}, \bibinfo {author} {\bibfnamefont {J.~C.~H.}\
  \bibnamefont {Spence}}, \bibinfo {author} {\bibfnamefont {A.}~\bibnamefont
  {Ourmazd}}, \bibinfo {author} {\bibfnamefont {P.}~\bibnamefont {Schwander}},
  \bibinfo {author} {\bibfnamefont {U.}~\bibnamefont {Weierstall}}, \bibinfo
  {author} {\bibfnamefont {M.}~\bibnamefont {Frank}}, \bibinfo {author}
  {\bibfnamefont {P.}~\bibnamefont {Fromme}}, \bibinfo {author} {\bibfnamefont
  {A.}~\bibnamefont {Barty}}, \bibinfo {author} {\bibfnamefont {H.~N.}\
  \bibnamefont {Chapman}}, \bibinfo {author} {\bibfnamefont {K.}~\bibnamefont
  {Moffat}}, \bibinfo {author} {\bibfnamefont {J.~J.}\ \bibnamefont {van
  Thor}}, and\ \bibinfo {author} {\bibfnamefont {M.}~\bibnamefont {Schmidt}},\
  }\bibfield  {title} {\enquote {\bibinfo {title} {Femtosecond structural
  dynamics drives the trans/cis isomerization in photoactive yellow protein},}\
  }\href {https://doi.org/10.1126/science.aad5081} {\bibfield  {journal}
  {\bibinfo  {journal} {Science}\ }\textbf {\bibinfo {volume} {352}},\ \bibinfo
  {pages} {725--729} (\bibinfo {year} {2016})}\BibitemShut {NoStop}%
\bibitem [{\citenamefont {K{\"u}pper}\ \emph {et~al.}(2014)\citenamefont
  {K{\"u}pper}, \citenamefont {Stern}, \citenamefont {Holmegaard},
  \citenamefont {Filsinger}, \citenamefont {Rouz\'{e}e}, \citenamefont
  {Rudenko}, \citenamefont {Johnsson}, \citenamefont {Martin}, \citenamefont
  {Adolph}, \citenamefont {Aquila}, \citenamefont {Bajt}, \citenamefont
  {Barty}, \citenamefont {Bostedt}, \citenamefont {Bozek}, \citenamefont
  {Caleman}, \citenamefont {Coffee}, \citenamefont {Coppola}, \citenamefont
  {Delmas}, \citenamefont {Epp}, \citenamefont {Erk}, \citenamefont {Foucar},
  \citenamefont {Gorkhover}, \citenamefont {Gumprecht}, \citenamefont
  {Hartmann}, \citenamefont {Hartmann}, \citenamefont {Hauser}, \citenamefont
  {Holl}, \citenamefont {H{\"o}mke}, \citenamefont {Kimmel}, \citenamefont
  {Krasniqi}, \citenamefont {K{\"u}hnel}, \citenamefont {Maurer}, \citenamefont
  {Messerschmidt}, \citenamefont {Moshammer}, \citenamefont {Reich},
  \citenamefont {Rudek}, \citenamefont {Santra}, \citenamefont {Schlichting},
  \citenamefont {Schmidt}, \citenamefont {Schorb}, \citenamefont {Schulz},
  \citenamefont {Soltau}, \citenamefont {Spence}, \citenamefont {Starodub},
  \citenamefont {Str{\"u}der}, \citenamefont {Th{\o}gersen}, \citenamefont
  {Vrakking}, \citenamefont {Weidenspointner}, \citenamefont {White},
  \citenamefont {Wunderer}, \citenamefont {Meijer}, \citenamefont {Ullrich},
  \citenamefont {Stapelfeldt}, \citenamefont {Rolles}, and\ \citenamefont
  {Chapman}}]{Kuepper:PRL112:083002}%
  \BibitemOpen
  \bibfield  {author} {\bibinfo {author} {\bibfnamefont {J.}~\bibnamefont
  {K{\"u}pper}}, \bibinfo {author} {\bibfnamefont {S.}~\bibnamefont {Stern}},
  \bibinfo {author} {\bibfnamefont {L.}~\bibnamefont {Holmegaard}}, \bibinfo
  {author} {\bibfnamefont {F.}~\bibnamefont {Filsinger}}, \bibinfo {author}
  {\bibfnamefont {A.}~\bibnamefont {Rouz\'{e}e}}, \bibinfo {author}
  {\bibfnamefont {A.}~\bibnamefont {Rudenko}}, \bibinfo {author} {\bibfnamefont
  {P.}~\bibnamefont {Johnsson}}, \bibinfo {author} {\bibfnamefont {A.~V.}\
  \bibnamefont {Martin}}, \bibinfo {author} {\bibfnamefont {M.}~\bibnamefont
  {Adolph}}, \bibinfo {author} {\bibfnamefont {A.}~\bibnamefont {Aquila}},
  \bibinfo {author} {\bibfnamefont {S.}~\bibnamefont {Bajt}}, \bibinfo {author}
  {\bibfnamefont {A.}~\bibnamefont {Barty}}, \bibinfo {author} {\bibfnamefont
  {C.}~\bibnamefont {Bostedt}}, \bibinfo {author} {\bibfnamefont
  {J.}~\bibnamefont {Bozek}}, \bibinfo {author} {\bibfnamefont
  {C.}~\bibnamefont {Caleman}}, \bibinfo {author} {\bibfnamefont
  {R.}~\bibnamefont {Coffee}}, \bibinfo {author} {\bibfnamefont
  {N.}~\bibnamefont {Coppola}}, \bibinfo {author} {\bibfnamefont
  {T.}~\bibnamefont {Delmas}}, \bibinfo {author} {\bibfnamefont
  {S.}~\bibnamefont {Epp}}, \bibinfo {author} {\bibfnamefont {B.}~\bibnamefont
  {Erk}}, \bibinfo {author} {\bibfnamefont {L.}~\bibnamefont {Foucar}},
  \bibinfo {author} {\bibfnamefont {T.}~\bibnamefont {Gorkhover}}, \bibinfo
  {author} {\bibfnamefont {L.}~\bibnamefont {Gumprecht}}, \bibinfo {author}
  {\bibfnamefont {A.}~\bibnamefont {Hartmann}}, \bibinfo {author}
  {\bibfnamefont {R.}~\bibnamefont {Hartmann}}, \bibinfo {author}
  {\bibfnamefont {G.}~\bibnamefont {Hauser}}, \bibinfo {author} {\bibfnamefont
  {P.}~\bibnamefont {Holl}}, \bibinfo {author} {\bibfnamefont {A.}~\bibnamefont
  {H{\"o}mke}}, \bibinfo {author} {\bibfnamefont {N.}~\bibnamefont {Kimmel}},
  \bibinfo {author} {\bibfnamefont {F.}~\bibnamefont {Krasniqi}}, \bibinfo
  {author} {\bibfnamefont {K.-U.}\ \bibnamefont {K{\"u}hnel}}, \bibinfo
  {author} {\bibfnamefont {J.}~\bibnamefont {Maurer}}, \bibinfo {author}
  {\bibfnamefont {M.}~\bibnamefont {Messerschmidt}}, \bibinfo {author}
  {\bibfnamefont {R.}~\bibnamefont {Moshammer}}, \bibinfo {author}
  {\bibfnamefont {C.}~\bibnamefont {Reich}}, \bibinfo {author} {\bibfnamefont
  {B.}~\bibnamefont {Rudek}}, \bibinfo {author} {\bibfnamefont
  {R.}~\bibnamefont {Santra}}, \bibinfo {author} {\bibfnamefont
  {I.}~\bibnamefont {Schlichting}}, \bibinfo {author} {\bibfnamefont
  {C.}~\bibnamefont {Schmidt}}, \bibinfo {author} {\bibfnamefont
  {S.}~\bibnamefont {Schorb}}, \bibinfo {author} {\bibfnamefont
  {J.}~\bibnamefont {Schulz}}, \bibinfo {author} {\bibfnamefont
  {H.}~\bibnamefont {Soltau}}, \bibinfo {author} {\bibfnamefont {J.~C.~H.}\
  \bibnamefont {Spence}}, \bibinfo {author} {\bibfnamefont {D.}~\bibnamefont
  {Starodub}}, \bibinfo {author} {\bibfnamefont {L.}~\bibnamefont
  {Str{\"u}der}}, \bibinfo {author} {\bibfnamefont {J.}~\bibnamefont
  {Th{\o}gersen}}, \bibinfo {author} {\bibfnamefont {M.~J.~J.}\ \bibnamefont
  {Vrakking}}, \bibinfo {author} {\bibfnamefont {G.}~\bibnamefont
  {Weidenspointner}}, \bibinfo {author} {\bibfnamefont {T.~A.}\ \bibnamefont
  {White}}, \bibinfo {author} {\bibfnamefont {C.}~\bibnamefont {Wunderer}},
  \bibinfo {author} {\bibfnamefont {G.}~\bibnamefont {Meijer}}, \bibinfo
  {author} {\bibfnamefont {J.}~\bibnamefont {Ullrich}}, \bibinfo {author}
  {\bibfnamefont {H.}~\bibnamefont {Stapelfeldt}}, \bibinfo {author}
  {\bibfnamefont {D.}~\bibnamefont {Rolles}}, and\ \bibinfo {author}
  {\bibfnamefont {H.~N.}\ \bibnamefont {Chapman}},\ }\bibfield  {title}
  {\enquote {\bibinfo {title} {X-ray diffraction from isolated and strongly
  aligned gas-phase molecules with a free-electron laser},}\ }\href
  {https://doi.org/10.1103/PhysRevLett.112.083002} {\bibfield  {journal}
  {\bibinfo  {journal} {Phys. Rev. Lett.}\ }\textbf {\bibinfo {volume} {112}},\
  \bibinfo {pages} {083002} (\bibinfo {year} {2014})},\ \Eprint
  {https://arxiv.org/abs/1307.4577} {arXiv:1307.4577 [physics]}\BibitemShut
  {NoStop}%
\bibitem [{\citenamefont {Stern}\ \emph {et~al.}(2014)\citenamefont {Stern},
  \citenamefont {Holmegaard}, \citenamefont {Filsinger}, \citenamefont
  {Rouz\'{e}e}, \citenamefont {Rudenko}, \citenamefont {Johnsson},
  \citenamefont {Martin}, \citenamefont {Barty}, \citenamefont {Bostedt},
  \citenamefont {Bozek}, \citenamefont {Coffee}, \citenamefont {Epp},
  \citenamefont {Erk}, \citenamefont {Foucar}, \citenamefont {Hartmann},
  \citenamefont {Kimmel}, \citenamefont {K{\"u}hnel}, \citenamefont {Maurer},
  \citenamefont {Messerschmidt}, \citenamefont {Rudek}, \citenamefont
  {Starodub}, \citenamefont {Th{\o}gersen}, \citenamefont {Weidenspointner},
  \citenamefont {White}, \citenamefont {Stapelfeldt}, \citenamefont {Rolles},
  \citenamefont {Chapman}, and\ \citenamefont {K{\"u}pper}}]{Stern:FD171:393}%
  \BibitemOpen
  \bibfield  {author} {\bibinfo {author} {\bibfnamefont {S.}~\bibnamefont
  {Stern}}, \bibinfo {author} {\bibfnamefont {L.}~\bibnamefont {Holmegaard}},
  \bibinfo {author} {\bibfnamefont {F.}~\bibnamefont {Filsinger}}, \bibinfo
  {author} {\bibfnamefont {A.}~\bibnamefont {Rouz\'{e}e}}, \bibinfo {author}
  {\bibfnamefont {A.}~\bibnamefont {Rudenko}}, \bibinfo {author} {\bibfnamefont
  {P.}~\bibnamefont {Johnsson}}, \bibinfo {author} {\bibfnamefont {A.~V.}\
  \bibnamefont {Martin}}, \bibinfo {author} {\bibfnamefont {A.}~\bibnamefont
  {Barty}}, \bibinfo {author} {\bibfnamefont {C.}~\bibnamefont {Bostedt}},
  \bibinfo {author} {\bibfnamefont {J.~D.}\ \bibnamefont {Bozek}}, \bibinfo
  {author} {\bibfnamefont {R.~N.}\ \bibnamefont {Coffee}}, \bibinfo {author}
  {\bibfnamefont {S.}~\bibnamefont {Epp}}, \bibinfo {author} {\bibfnamefont
  {B.}~\bibnamefont {Erk}}, \bibinfo {author} {\bibfnamefont {L.}~\bibnamefont
  {Foucar}}, \bibinfo {author} {\bibfnamefont {R.}~\bibnamefont {Hartmann}},
  \bibinfo {author} {\bibfnamefont {N.}~\bibnamefont {Kimmel}}, \bibinfo
  {author} {\bibfnamefont {K.-U.}\ \bibnamefont {K{\"u}hnel}}, \bibinfo
  {author} {\bibfnamefont {J.}~\bibnamefont {Maurer}}, \bibinfo {author}
  {\bibfnamefont {M.}~\bibnamefont {Messerschmidt}}, \bibinfo {author}
  {\bibfnamefont {B.}~\bibnamefont {Rudek}}, \bibinfo {author} {\bibfnamefont
  {D.~G.}\ \bibnamefont {Starodub}}, \bibinfo {author} {\bibfnamefont
  {J.}~\bibnamefont {Th{\o}gersen}}, \bibinfo {author} {\bibfnamefont
  {G.}~\bibnamefont {Weidenspointner}}, \bibinfo {author} {\bibfnamefont
  {T.~A.}\ \bibnamefont {White}}, \bibinfo {author} {\bibfnamefont
  {H.}~\bibnamefont {Stapelfeldt}}, \bibinfo {author} {\bibfnamefont
  {D.}~\bibnamefont {Rolles}}, \bibinfo {author} {\bibfnamefont {H.~N.}\
  \bibnamefont {Chapman}}, and\ \bibinfo {author} {\bibfnamefont
  {J.}~\bibnamefont {K{\"u}pper}},\ }\bibfield  {title} {\enquote {\bibinfo
  {title} {Toward atomic resolution diffractive imaging of isolated molecules
  with x-ray free-electron lasers},}\ }\href
  {https://doi.org/10.1039/c4fd00028e} {\bibfield  {journal} {\bibinfo
  {journal} {Faraday Disc.}\ }\textbf {\bibinfo {volume} {171}},\ \bibinfo
  {pages} {393} (\bibinfo {year} {2014})},\ \Eprint
  {https://arxiv.org/abs/1403.2553} {arXiv:1403.2553 [physics]}\BibitemShut
  {NoStop}%
\bibitem [{\citenamefont {Zewail}(2006)}]{Zewail:ARPC57:65}%
  \BibitemOpen
  \bibfield  {author} {\bibinfo {author} {\bibfnamefont {A.~H.}\ \bibnamefont
  {Zewail}},\ }\bibfield  {title} {\enquote {\bibinfo {title} {{4D} ultrafast
  electron diffraction, crystallography, and microscopy},}\ }\href@noop {}
  {\bibfield  {journal} {\bibinfo  {journal} {Annu. Rev. Phys. Chem.}\ }\textbf
  {\bibinfo {volume} {57}},\ \bibinfo {pages} {65--103} (\bibinfo {year}
  {2006})}\BibitemShut {NoStop}%
\bibitem [{\citenamefont {Ischenko}, \citenamefont {Weber}, and\ \citenamefont
  {Miller}(2017)}]{Ischenko:CR117:11066}%
  \BibitemOpen
  \bibfield  {author} {\bibinfo {author} {\bibfnamefont {A.~A.}\ \bibnamefont
  {Ischenko}}, \bibinfo {author} {\bibfnamefont {P.~M.}\ \bibnamefont
  {Weber}}, and\ \bibinfo {author} {\bibfnamefont {R.~J.~D.}\ \bibnamefont
  {Miller}},\ }\bibfield  {title} {\enquote {\bibinfo {title} {Capturing
  chemistry in action with electrons: Realization of atomically resolved
  reaction dynamics},}\ }\href {https://doi.org/10.1021/acs.chemrev.6b00770}
  {\bibfield  {journal} {\bibinfo  {journal} {Chem. Rev.}\ }\textbf {\bibinfo
  {volume} {117}},\ \bibinfo {pages} {11066--11124} (\bibinfo {year}
  {2017})}\BibitemShut {NoStop}%
\bibitem [{\citenamefont {Blaga}\ \emph {et~al.}(2012)\citenamefont {Blaga},
  \citenamefont {Xu}, \citenamefont {DiChiara}, \citenamefont {Sistrunk},
  \citenamefont {Zhang}, \citenamefont {Agostini}, \citenamefont {Miller},
  \citenamefont {DiMauro}, and\ \citenamefont {Lin}}]{Blaga:Nature483:194}%
  \BibitemOpen
  \bibfield  {author} {\bibinfo {author} {\bibfnamefont {C.~I.}\ \bibnamefont
  {Blaga}}, \bibinfo {author} {\bibfnamefont {J.}~\bibnamefont {Xu}}, \bibinfo
  {author} {\bibfnamefont {A.~D.}\ \bibnamefont {DiChiara}}, \bibinfo {author}
  {\bibfnamefont {E.}~\bibnamefont {Sistrunk}}, \bibinfo {author}
  {\bibfnamefont {K.}~\bibnamefont {Zhang}}, \bibinfo {author} {\bibfnamefont
  {P.}~\bibnamefont {Agostini}}, \bibinfo {author} {\bibfnamefont {T.~A.}\
  \bibnamefont {Miller}}, \bibinfo {author} {\bibfnamefont {L.~F.}\
  \bibnamefont {DiMauro}}, and\ \bibinfo {author} {\bibfnamefont {C.~D.}\
  \bibnamefont {Lin}},\ }\bibfield  {title} {\enquote {\bibinfo {title}
  {Imaging ultrafast molecular dynamics with laser-induced electron
  diffraction},}\ }\href {https://doi.org/10.1038/nature10820} {\bibfield
  {journal} {\bibinfo  {journal} {Nature}\ }\textbf {\bibinfo {volume} {483}},\
  \bibinfo {pages} {194--197} (\bibinfo {year} {2012})}\BibitemShut {NoStop}%
\bibitem [{\citenamefont {Wolter}\ \emph {et~al.}(2016)\citenamefont {Wolter},
  \citenamefont {Pullen}, \citenamefont {Le}, \citenamefont {Baudisch},
  \citenamefont {Doblhoff-Dier}, \citenamefont {Senftleben}, \citenamefont
  {Hemmer}, \citenamefont {Schroter}, \citenamefont {Ullrich}, \citenamefont
  {Pfeifer}, \citenamefont {Moshammer}, \citenamefont {Gr{\"a}fe},
  \citenamefont {Vendrell}, \citenamefont {Lin}, and\ \citenamefont
  {Biegert}}]{Wolter:Science354:308}%
  \BibitemOpen
  \bibfield  {author} {\bibinfo {author} {\bibfnamefont {B.}~\bibnamefont
  {Wolter}}, \bibinfo {author} {\bibfnamefont {M.~G.}\ \bibnamefont {Pullen}},
  \bibinfo {author} {\bibfnamefont {A.~T.}\ \bibnamefont {Le}}, \bibinfo
  {author} {\bibfnamefont {M.}~\bibnamefont {Baudisch}}, \bibinfo {author}
  {\bibfnamefont {K.}~\bibnamefont {Doblhoff-Dier}}, \bibinfo {author}
  {\bibfnamefont {A.}~\bibnamefont {Senftleben}}, \bibinfo {author}
  {\bibfnamefont {M.}~\bibnamefont {Hemmer}}, \bibinfo {author} {\bibfnamefont
  {C.~D.}\ \bibnamefont {Schroter}}, \bibinfo {author} {\bibfnamefont
  {J.}~\bibnamefont {Ullrich}}, \bibinfo {author} {\bibfnamefont
  {T.}~\bibnamefont {Pfeifer}}, \bibinfo {author} {\bibfnamefont
  {R.}~\bibnamefont {Moshammer}}, \bibinfo {author} {\bibfnamefont
  {S.}~\bibnamefont {Gr{\"a}fe}}, \bibinfo {author} {\bibfnamefont
  {O.}~\bibnamefont {Vendrell}}, \bibinfo {author} {\bibfnamefont {C.~D.}\
  \bibnamefont {Lin}}, and\ \bibinfo {author} {\bibfnamefont {J.}~\bibnamefont
  {Biegert}},\ }\bibfield  {title} {\enquote {\bibinfo {title} {Ultrafast
  electron diffraction imaging of bond breaking in di-ionized acetylene},}\
  }\href {https://doi.org/10.1126/science.aah3429} {\bibfield  {journal}
  {\bibinfo  {journal} {Science}\ }\textbf {\bibinfo {volume} {354}},\ \bibinfo
  {pages} {308--312} (\bibinfo {year} {2016})}\BibitemShut {NoStop}%
\bibitem [{\citenamefont {Stapelfeldt}, \citenamefont {Constant}, and\
  \citenamefont {Corkum}(1995)}]{Stapelfeldt:PRL74:3780}%
  \BibitemOpen
  \bibfield  {author} {\bibinfo {author} {\bibfnamefont {H.}~\bibnamefont
  {Stapelfeldt}}, \bibinfo {author} {\bibfnamefont {E.}~\bibnamefont
  {Constant}}, and\ \bibinfo {author} {\bibfnamefont {P.~B.}\ \bibnamefont
  {Corkum}},\ }\bibfield  {title} {\enquote {\bibinfo {title} {Wave-packet
  structure and dynamics measured by {C}oulomb explosion},}\ }\href
  {https://doi.org/10.1103/PhysRevA.58.426} {\bibfield  {journal} {\bibinfo
  {journal} {Phys. Rev. Lett.}\ }\textbf {\bibinfo {volume} {74}},\ \bibinfo
  {pages} {3780--3783} (\bibinfo {year} {1995})}\BibitemShut {NoStop}%
\bibitem [{\citenamefont {Pitzer}\ \emph {et~al.}(2013)\citenamefont {Pitzer},
  \citenamefont {Kunitski}, \citenamefont {Johnson}, \citenamefont {Jahnke},
  \citenamefont {Sann}, \citenamefont {Sturm}, \citenamefont {Schmidt},
  \citenamefont {Schmidt-Böcking}, \citenamefont {Dörner}, \citenamefont
  {Stohner}, \citenamefont {Kiedrowski}, \citenamefont {Reggelin},
  \citenamefont {Marquardt}, \citenamefont {Schießer}, \citenamefont
  {Berger}, and\ \citenamefont {Schöffler}}]{Pitzer:Science341:1096}%
  \BibitemOpen
  \bibfield  {author} {\bibinfo {author} {\bibfnamefont {M.}~\bibnamefont
  {Pitzer}}, \bibinfo {author} {\bibfnamefont {M.}~\bibnamefont {Kunitski}},
  \bibinfo {author} {\bibfnamefont {A.~S.}\ \bibnamefont {Johnson}}, \bibinfo
  {author} {\bibfnamefont {T.}~\bibnamefont {Jahnke}}, \bibinfo {author}
  {\bibfnamefont {H.}~\bibnamefont {Sann}}, \bibinfo {author} {\bibfnamefont
  {F.}~\bibnamefont {Sturm}}, \bibinfo {author} {\bibfnamefont {L.~P.~H.}\
  \bibnamefont {Schmidt}}, \bibinfo {author} {\bibfnamefont {H.}~\bibnamefont
  {Schmidt-Böcking}}, \bibinfo {author} {\bibfnamefont {R.}~\bibnamefont
  {Dörner}}, \bibinfo {author} {\bibfnamefont {J.}~\bibnamefont {Stohner}},
  \bibinfo {author} {\bibfnamefont {J.}~\bibnamefont {Kiedrowski}}, \bibinfo
  {author} {\bibfnamefont {M.}~\bibnamefont {Reggelin}}, \bibinfo {author}
  {\bibfnamefont {S.}~\bibnamefont {Marquardt}}, \bibinfo {author}
  {\bibfnamefont {A.}~\bibnamefont {Schießer}}, \bibinfo {author}
  {\bibfnamefont {R.}~\bibnamefont {Berger}}, and\ \bibinfo {author}
  {\bibfnamefont {M.~S.}\ \bibnamefont {Schöffler}},\ }\bibfield  {title}
  {\enquote {\bibinfo {title} {Direct determination of absolute molecular
  stereochemistry in gas phase by {C}oulomb explosion imaging},}\ }\href
  {https://doi.org/10.1126/science.1240362} {\bibfield  {journal} {\bibinfo
  {journal} {Science}\ }\textbf {\bibinfo {volume} {341}},\ \bibinfo {pages}
  {1096--1100} (\bibinfo {year} {2013})}\BibitemShut {NoStop}%
\bibitem [{\citenamefont {Nagaya}\ \emph {et~al.}(2016)\citenamefont {Nagaya},
  \citenamefont {Motomura}, \citenamefont {Kukk}, \citenamefont {Takahashi},
  \citenamefont {Yamazaki}, \citenamefont {Ohmura}, \citenamefont {Fukuzawa},
  \citenamefont {Wada}, \citenamefont {Mondal}, \citenamefont {Tachibana},
  \citenamefont {Ito}, \citenamefont {Koga}, \citenamefont {Sakai},
  \citenamefont {Matsunami}, \citenamefont {Nakamura}, \citenamefont {Kanno},
  \citenamefont {Rudenko}, \citenamefont {Nicolas}, \citenamefont {Liu},
  \citenamefont {Miron}, \citenamefont {Zhang}, \citenamefont {Jiang},
  \citenamefont {Chen}, \citenamefont {Anand}, \citenamefont {Kim},
  \citenamefont {Tono}, \citenamefont {Yabashi}, \citenamefont {Yao},
  \citenamefont {Kono}, and\ \citenamefont {Ueda}}]{Nagaya:FD194:537}%
  \BibitemOpen
  \bibfield  {author} {\bibinfo {author} {\bibfnamefont {K.}~\bibnamefont
  {Nagaya}}, \bibinfo {author} {\bibfnamefont {K.}~\bibnamefont {Motomura}},
  \bibinfo {author} {\bibfnamefont {E.}~\bibnamefont {Kukk}}, \bibinfo {author}
  {\bibfnamefont {Y.}~\bibnamefont {Takahashi}}, \bibinfo {author}
  {\bibfnamefont {K.}~\bibnamefont {Yamazaki}}, \bibinfo {author}
  {\bibfnamefont {S.}~\bibnamefont {Ohmura}}, \bibinfo {author} {\bibfnamefont
  {H.}~\bibnamefont {Fukuzawa}}, \bibinfo {author} {\bibfnamefont
  {S.}~\bibnamefont {Wada}}, \bibinfo {author} {\bibfnamefont {S.}~\bibnamefont
  {Mondal}}, \bibinfo {author} {\bibfnamefont {T.}~\bibnamefont {Tachibana}},
  \bibinfo {author} {\bibfnamefont {Y.}~\bibnamefont {Ito}}, \bibinfo {author}
  {\bibfnamefont {R.}~\bibnamefont {Koga}}, \bibinfo {author} {\bibfnamefont
  {T.}~\bibnamefont {Sakai}}, \bibinfo {author} {\bibfnamefont
  {K.}~\bibnamefont {Matsunami}}, \bibinfo {author} {\bibfnamefont
  {K.}~\bibnamefont {Nakamura}}, \bibinfo {author} {\bibfnamefont
  {M.}~\bibnamefont {Kanno}}, \bibinfo {author} {\bibfnamefont
  {A.}~\bibnamefont {Rudenko}}, \bibinfo {author} {\bibfnamefont
  {C.}~\bibnamefont {Nicolas}}, \bibinfo {author} {\bibfnamefont {X.~J.}\
  \bibnamefont {Liu}}, \bibinfo {author} {\bibfnamefont {C.}~\bibnamefont
  {Miron}}, \bibinfo {author} {\bibfnamefont {Y.}~\bibnamefont {Zhang}},
  \bibinfo {author} {\bibfnamefont {Y.}~\bibnamefont {Jiang}}, \bibinfo
  {author} {\bibfnamefont {J.}~\bibnamefont {Chen}}, \bibinfo {author}
  {\bibfnamefont {M.}~\bibnamefont {Anand}}, \bibinfo {author} {\bibfnamefont
  {D.~E.}\ \bibnamefont {Kim}}, \bibinfo {author} {\bibfnamefont
  {K.}~\bibnamefont {Tono}}, \bibinfo {author} {\bibfnamefont {M.}~\bibnamefont
  {Yabashi}}, \bibinfo {author} {\bibfnamefont {M.}~\bibnamefont {Yao}},
  \bibinfo {author} {\bibfnamefont {H.}~\bibnamefont {Kono}}, and\ \bibinfo
  {author} {\bibfnamefont {K.}~\bibnamefont {Ueda}},\ }\bibfield  {title}
  {\enquote {\bibinfo {title} {Femtosecond charge and molecular dynamics of
  {I}-containing organic molecules induced by intense x-ray free-electron laser
  pulses},}\ }\href {https://doi.org/10.1039/c6fd00085a} {\bibfield  {journal}
  {\bibinfo  {journal} {Faraday Disc.}\ }\textbf {\bibinfo {volume} {194}},\
  \bibinfo {pages} {537--562} (\bibinfo {year} {2016})}\BibitemShut {NoStop}%
\bibitem [{\citenamefont {Itatani}\ \emph {et~al.}(2004)\citenamefont
  {Itatani}, \citenamefont {Levesque}, \citenamefont {Zeidler}, \citenamefont
  {Niikura}, \citenamefont {P\'{e}pin}, \citenamefont {Kieffer}, \citenamefont
  {Corkum}, and\ \citenamefont {Villeneuve}}]{Itatani:Nature432:867}%
  \BibitemOpen
  \bibfield  {author} {\bibinfo {author} {\bibfnamefont {J.}~\bibnamefont
  {Itatani}}, \bibinfo {author} {\bibfnamefont {J.}~\bibnamefont {Levesque}},
  \bibinfo {author} {\bibfnamefont {D.}~\bibnamefont {Zeidler}}, \bibinfo
  {author} {\bibfnamefont {H.}~\bibnamefont {Niikura}}, \bibinfo {author}
  {\bibfnamefont {H.}~\bibnamefont {P\'{e}pin}}, \bibinfo {author}
  {\bibfnamefont {J.~C.}\ \bibnamefont {Kieffer}}, \bibinfo {author}
  {\bibfnamefont {P.~B.}\ \bibnamefont {Corkum}}, and\ \bibinfo {author}
  {\bibfnamefont {D.~M.}\ \bibnamefont {Villeneuve}},\ }\bibfield  {title}
  {\enquote {\bibinfo {title} {Tomographic imaging of molecular orbitals},}\
  }\href {https://doi.org/10.1038/nature03183} {\bibfield  {journal} {\bibinfo
  {journal} {Nature}\ }\textbf {\bibinfo {volume} {432}},\ \bibinfo {pages}
  {867--871} (\bibinfo {year} {2004})}\BibitemShut {NoStop}%
\bibitem [{\citenamefont {Vozzi}\ \emph {et~al.}(2011)\citenamefont {Vozzi},
  \citenamefont {Negro}, \citenamefont {Calegari}, \citenamefont {Sansone},
  \citenamefont {Nisoli}, \citenamefont {De~Silvestri}, and\ \citenamefont
  {Stagira}}]{Vozzi:NatPhys7:822}%
  \BibitemOpen
  \bibfield  {author} {\bibinfo {author} {\bibfnamefont {C.}~\bibnamefont
  {Vozzi}}, \bibinfo {author} {\bibfnamefont {M.}~\bibnamefont {Negro}},
  \bibinfo {author} {\bibfnamefont {F.}~\bibnamefont {Calegari}}, \bibinfo
  {author} {\bibfnamefont {G.}~\bibnamefont {Sansone}}, \bibinfo {author}
  {\bibfnamefont {M.}~\bibnamefont {Nisoli}}, \bibinfo {author} {\bibfnamefont
  {S.}~\bibnamefont {De~Silvestri}}, and\ \bibinfo {author} {\bibfnamefont
  {S.}~\bibnamefont {Stagira}},\ }\bibfield  {title} {\enquote {\bibinfo
  {title} {Generalized molecular orbital tomography},}\ }\href
  {https://doi.org/10.1038/nphys2029} {\bibfield  {journal} {\bibinfo
  {journal} {Nat. Phys.}\ }\textbf {\bibinfo {volume} {7}},\ \bibinfo {pages}
  {822--826} (\bibinfo {year} {2011})}\BibitemShut {NoStop}%
\bibitem [{\citenamefont {Wörner}\ \emph {et~al.}(2010)\citenamefont
  {Wörner}, \citenamefont {Bertrand}, \citenamefont {Kartashov}, \citenamefont
  {Corkum}, and\ \citenamefont {Villeneuve}}]{Woerner:Nature466:604}%
  \BibitemOpen
  \bibfield  {author} {\bibinfo {author} {\bibfnamefont {H.~J.}\ \bibnamefont
  {Wörner}}, \bibinfo {author} {\bibfnamefont {J.~B.}\ \bibnamefont
  {Bertrand}}, \bibinfo {author} {\bibfnamefont {D.~V.}\ \bibnamefont
  {Kartashov}}, \bibinfo {author} {\bibfnamefont {P.~B.}\ \bibnamefont
  {Corkum}}, and\ \bibinfo {author} {\bibfnamefont {D.~M.}\ \bibnamefont
  {Villeneuve}},\ }\bibfield  {title} {\enquote {\bibinfo {title} {Following a
  chemical reaction using high-harmonic interferometry},}\ }\href
  {https://doi.org/10.1038/nature09185} {\bibfield  {journal} {\bibinfo
  {journal} {Nature}\ }\textbf {\bibinfo {volume} {466}},\ \bibinfo {pages}
  {604--607} (\bibinfo {year} {2010})}\BibitemShut {NoStop}%
\bibitem [{\citenamefont {Baumert}\ \emph {et~al.}(1990)\citenamefont
  {Baumert}, \citenamefont {Bühler}, \citenamefont {Thalweiser}, and\
  \citenamefont {Gerber}}]{Baumert:PRL64:733}%
  \BibitemOpen
  \bibfield  {author} {\bibinfo {author} {\bibfnamefont {T.}~\bibnamefont
  {Baumert}}, \bibinfo {author} {\bibfnamefont {B.}~\bibnamefont {Bühler}},
  \bibinfo {author} {\bibfnamefont {R.}~\bibnamefont {Thalweiser}}, and\
  \bibinfo {author} {\bibfnamefont {G.}~\bibnamefont {Gerber}},\ }\bibfield
  {title} {\enquote {\bibinfo {title} {Femtosecond spectroscopy of molecular
  autoionization and fragmentation},}\ }\href
  {https://doi.org/10.1103/physrevlett.64.733} {\bibfield  {journal} {\bibinfo
  {journal} {Phys. Rev. Lett.}\ }\textbf {\bibinfo {volume} {64}},\ \bibinfo
  {pages} {733--736} (\bibinfo {year} {1990})}\BibitemShut {NoStop}%
\bibitem [{\citenamefont {Stolow} and\ \citenamefont
  {Underwood}(2008)}]{Stolow:ACP139:497}%
  \BibitemOpen
  \bibfield  {author} {\bibinfo {author} {\bibfnamefont {A.}~\bibnamefont
  {Stolow}} and\ \bibinfo {author} {\bibfnamefont {J.~G.}\ \bibnamefont
  {Underwood}},\ }\bibfield  {title} {\enquote {\bibinfo {title} {Time-resolved
  photoelectron spectroscopy of nonadiabatic dynamics in polyatomic
  molecules},}\ }\href {https://doi.org/10.1002/9780470259498.ch6} {\bibfield
  {journal} {\bibinfo  {journal} {Adv. Chem. Phys.}\ }\textbf {\bibinfo
  {volume} {139}},\ \bibinfo {pages} {497–584} (\bibinfo {year}
  {2008})}\BibitemShut {NoStop}%
\bibitem [{\citenamefont {Boll}\ \emph {et~al.}(2014)\citenamefont {Boll},
  \citenamefont {Rouz{\'e}e}, \citenamefont {Adolph}, \citenamefont {Anielski},
  \citenamefont {Aquila}, \citenamefont {Bari}, \citenamefont {Bomme},
  \citenamefont {Bostedt}, \citenamefont {Bozek}, \citenamefont {Chapman},
  \citenamefont {Christensen}, \citenamefont {Coffee}, \citenamefont {Coppola},
  \citenamefont {De}, \citenamefont {Decleva}, \citenamefont {Epp},
  \citenamefont {Erk}, \citenamefont {Filsinger}, \citenamefont {Foucar},
  \citenamefont {Gorkhover}, \citenamefont {Gumprecht}, \citenamefont
  {H{\"o}mke}, \citenamefont {Holmegaard}, \citenamefont {Johnsson},
  \citenamefont {Kienitz}, \citenamefont {Kierspel}, \citenamefont {Krasniqi},
  \citenamefont {K{\"u}hnel}, \citenamefont {Maurer}, \citenamefont
  {Messerschmidt}, \citenamefont {Moshammer}, \citenamefont {M{\"u}ller},
  \citenamefont {Rudek}, \citenamefont {Savelyev}, \citenamefont {Schlichting},
  \citenamefont {Schmidt}, \citenamefont {Scholz}, \citenamefont {Schorb},
  \citenamefont {Schulz}, \citenamefont {Seltmann}, \citenamefont {Stener},
  \citenamefont {Stern}, \citenamefont {Techert}, \citenamefont {Th{\o}gersen},
  \citenamefont {Trippel}, \citenamefont {Viefhaus}, \citenamefont {Vrakking},
  \citenamefont {Stapelfeldt}, \citenamefont {K{\"u}pper}, \citenamefont
  {Ullrich}, \citenamefont {Rudenko}, and\ \citenamefont
  {Rolles}}]{Boll:FD171:57}%
  \BibitemOpen
  \bibfield  {author} {\bibinfo {author} {\bibfnamefont {R.}~\bibnamefont
  {Boll}}, \bibinfo {author} {\bibfnamefont {A.}~\bibnamefont {Rouz{\'e}e}},
  \bibinfo {author} {\bibfnamefont {M.}~\bibnamefont {Adolph}}, \bibinfo
  {author} {\bibfnamefont {D.}~\bibnamefont {Anielski}}, \bibinfo {author}
  {\bibfnamefont {A.}~\bibnamefont {Aquila}}, \bibinfo {author} {\bibfnamefont
  {S.}~\bibnamefont {Bari}}, \bibinfo {author} {\bibfnamefont {C.}~\bibnamefont
  {Bomme}}, \bibinfo {author} {\bibfnamefont {C.}~\bibnamefont {Bostedt}},
  \bibinfo {author} {\bibfnamefont {J.~D.}\ \bibnamefont {Bozek}}, \bibinfo
  {author} {\bibfnamefont {H.~N.}\ \bibnamefont {Chapman}}, \bibinfo {author}
  {\bibfnamefont {L.}~\bibnamefont {Christensen}}, \bibinfo {author}
  {\bibfnamefont {R.}~\bibnamefont {Coffee}}, \bibinfo {author} {\bibfnamefont
  {N.}~\bibnamefont {Coppola}}, \bibinfo {author} {\bibfnamefont
  {S.}~\bibnamefont {De}}, \bibinfo {author} {\bibfnamefont {P.}~\bibnamefont
  {Decleva}}, \bibinfo {author} {\bibfnamefont {S.~W.}\ \bibnamefont {Epp}},
  \bibinfo {author} {\bibfnamefont {B.}~\bibnamefont {Erk}}, \bibinfo {author}
  {\bibfnamefont {F.}~\bibnamefont {Filsinger}}, \bibinfo {author}
  {\bibfnamefont {L.}~\bibnamefont {Foucar}}, \bibinfo {author} {\bibfnamefont
  {T.}~\bibnamefont {Gorkhover}}, \bibinfo {author} {\bibfnamefont
  {L.}~\bibnamefont {Gumprecht}}, \bibinfo {author} {\bibfnamefont
  {A.}~\bibnamefont {H{\"o}mke}}, \bibinfo {author} {\bibfnamefont
  {L.}~\bibnamefont {Holmegaard}}, \bibinfo {author} {\bibfnamefont
  {P.}~\bibnamefont {Johnsson}}, \bibinfo {author} {\bibfnamefont {J.~S.}\
  \bibnamefont {Kienitz}}, \bibinfo {author} {\bibfnamefont {T.}~\bibnamefont
  {Kierspel}}, \bibinfo {author} {\bibfnamefont {F.}~\bibnamefont {Krasniqi}},
  \bibinfo {author} {\bibfnamefont {K.-U.}\ \bibnamefont {K{\"u}hnel}},
  \bibinfo {author} {\bibfnamefont {J.}~\bibnamefont {Maurer}}, \bibinfo
  {author} {\bibfnamefont {M.}~\bibnamefont {Messerschmidt}}, \bibinfo {author}
  {\bibfnamefont {R.}~\bibnamefont {Moshammer}}, \bibinfo {author}
  {\bibfnamefont {N.~L.~M.}\ \bibnamefont {M{\"u}ller}}, \bibinfo {author}
  {\bibfnamefont {B.}~\bibnamefont {Rudek}}, \bibinfo {author} {\bibfnamefont
  {E.}~\bibnamefont {Savelyev}}, \bibinfo {author} {\bibfnamefont
  {I.}~\bibnamefont {Schlichting}}, \bibinfo {author} {\bibfnamefont
  {C.}~\bibnamefont {Schmidt}}, \bibinfo {author} {\bibfnamefont
  {F.}~\bibnamefont {Scholz}}, \bibinfo {author} {\bibfnamefont
  {S.}~\bibnamefont {Schorb}}, \bibinfo {author} {\bibfnamefont
  {J.}~\bibnamefont {Schulz}}, \bibinfo {author} {\bibfnamefont
  {J.}~\bibnamefont {Seltmann}}, \bibinfo {author} {\bibfnamefont
  {M.}~\bibnamefont {Stener}}, \bibinfo {author} {\bibfnamefont
  {S.}~\bibnamefont {Stern}}, \bibinfo {author} {\bibfnamefont
  {S.}~\bibnamefont {Techert}}, \bibinfo {author} {\bibfnamefont
  {J.}~\bibnamefont {Th{\o}gersen}}, \bibinfo {author} {\bibfnamefont
  {S.}~\bibnamefont {Trippel}}, \bibinfo {author} {\bibfnamefont
  {J.}~\bibnamefont {Viefhaus}}, \bibinfo {author} {\bibfnamefont
  {M.}~\bibnamefont {Vrakking}}, \bibinfo {author} {\bibfnamefont
  {H.}~\bibnamefont {Stapelfeldt}}, \bibinfo {author} {\bibfnamefont
  {J.}~\bibnamefont {K{\"u}pper}}, \bibinfo {author} {\bibfnamefont
  {J.}~\bibnamefont {Ullrich}}, \bibinfo {author} {\bibfnamefont
  {A.}~\bibnamefont {Rudenko}}, and\ \bibinfo {author} {\bibfnamefont
  {D.}~\bibnamefont {Rolles}},\ }\bibfield  {title} {\enquote {\bibinfo {title}
  {Imaging molecular structure through femtosecond photoelectron diffraction on
  aligned and oriented gas-phase molecules},}\ }\href
  {https://doi.org/10.1039/c4fd00037d} {\bibfield  {journal} {\bibinfo
  {journal} {Faraday Disc.}\ }\textbf {\bibinfo {volume} {171}},\ \bibinfo
  {pages} {57 -- 80} (\bibinfo {year} {2014})},\ \Eprint
  {https://arxiv.org/abs/1407.7782} {arXiv:1407.7782 [physics]}\BibitemShut
  {NoStop}%
\bibitem [{\citenamefont {Williamson}\ \emph {et~al.}(1997)\citenamefont
  {Williamson}, \citenamefont {Cao}, \citenamefont {Ihee}, \citenamefont
  {Frey}, and\ \citenamefont {Zewail}}]{Williamson:Nature386:159}%
  \BibitemOpen
  \bibfield  {author} {\bibinfo {author} {\bibfnamefont {J.~C.}\ \bibnamefont
  {Williamson}}, \bibinfo {author} {\bibfnamefont {J.~M.}\ \bibnamefont {Cao}},
  \bibinfo {author} {\bibfnamefont {H.}~\bibnamefont {Ihee}}, \bibinfo {author}
  {\bibfnamefont {H.}~\bibnamefont {Frey}}, and\ \bibinfo {author}
  {\bibfnamefont {A.~H.}\ \bibnamefont {Zewail}},\ }\bibfield  {title}
  {\enquote {\bibinfo {title} {Clocking transient chemical changes by ultrafast
  electron diffraction},}\ }\href {https://doi.org/10.1038/386159a0} {\bibfield
   {journal} {\bibinfo  {journal} {Nature}\ }\textbf {\bibinfo {volume}
  {386}},\ \bibinfo {pages} {159--162} (\bibinfo {year} {1997})}\BibitemShut
  {NoStop}%
\bibitem [{\citenamefont {Yang}\ \emph {et~al.}(2018)\citenamefont {Yang},
  \citenamefont {Zhu}, \citenamefont {Wolf}, \citenamefont {Li}, \citenamefont
  {Nunes}, \citenamefont {Coffee}, \citenamefont {Cryan}, \citenamefont
  {G{\"u}hr}, \citenamefont {Hegazy}, \citenamefont {Heinz}, \citenamefont
  {Jobe}, \citenamefont {Li}, \citenamefont {Shen}, \citenamefont {Veccione},
  \citenamefont {Weathersby}, \citenamefont {Wilkin}, \citenamefont {Yoneda},
  \citenamefont {Zheng}, \citenamefont {Mart{\'\i}nez}, \citenamefont
  {Centurion}, and\ \citenamefont {Wang}}]{Yang:Science361:64}%
  \BibitemOpen
  \bibfield  {author} {\bibinfo {author} {\bibfnamefont {J.}~\bibnamefont
  {Yang}}, \bibinfo {author} {\bibfnamefont {X.}~\bibnamefont {Zhu}}, \bibinfo
  {author} {\bibfnamefont {T.~J.~A.}\ \bibnamefont {Wolf}}, \bibinfo {author}
  {\bibfnamefont {Z.}~\bibnamefont {Li}}, \bibinfo {author} {\bibfnamefont
  {J.~P.~F.}\ \bibnamefont {Nunes}}, \bibinfo {author} {\bibfnamefont
  {R.}~\bibnamefont {Coffee}}, \bibinfo {author} {\bibfnamefont {J.~P.}\
  \bibnamefont {Cryan}}, \bibinfo {author} {\bibfnamefont {M.}~\bibnamefont
  {G{\"u}hr}}, \bibinfo {author} {\bibfnamefont {K.}~\bibnamefont {Hegazy}},
  \bibinfo {author} {\bibfnamefont {T.~F.}\ \bibnamefont {Heinz}}, \bibinfo
  {author} {\bibfnamefont {K.}~\bibnamefont {Jobe}}, \bibinfo {author}
  {\bibfnamefont {R.}~\bibnamefont {Li}}, \bibinfo {author} {\bibfnamefont
  {X.}~\bibnamefont {Shen}}, \bibinfo {author} {\bibfnamefont {T.}~\bibnamefont
  {Veccione}}, \bibinfo {author} {\bibfnamefont {S.}~\bibnamefont
  {Weathersby}}, \bibinfo {author} {\bibfnamefont {K.~J.}\ \bibnamefont
  {Wilkin}}, \bibinfo {author} {\bibfnamefont {C.}~\bibnamefont {Yoneda}},
  \bibinfo {author} {\bibfnamefont {Q.}~\bibnamefont {Zheng}}, \bibinfo
  {author} {\bibfnamefont {T.~J.}\ \bibnamefont {Mart{\'\i}nez}}, \bibinfo
  {author} {\bibfnamefont {M.}~\bibnamefont {Centurion}}, and\ \bibinfo
  {author} {\bibfnamefont {X.}~\bibnamefont {Wang}},\ }\bibfield  {title}
  {\enquote {\bibinfo {title} {Imaging {CF$_3$I} conical intersection and
  photodissociation dynamics with ultrafast electron diffraction},}\ }\href
  {https://doi.org/10.1126/science.aat0049} {\bibfield  {journal} {\bibinfo
  {journal} {Science}\ }\textbf {\bibinfo {volume} {361}},\ \bibinfo {pages}
  {64--67} (\bibinfo {year} {2018})}\BibitemShut {NoStop}%
\bibitem [{\citenamefont {Wolf}\ \emph {et~al.}(2019)\citenamefont {Wolf},
  \citenamefont {Sanchez}, \citenamefont {Yang}, \citenamefont {Parrish},
  \citenamefont {Nunes}, \citenamefont {Centurion}, \citenamefont {Coffee},
  \citenamefont {Cryan}, \citenamefont {G{\"u}hr}, \citenamefont {Hegazy},
  \citenamefont {Kirrander}, \citenamefont {Li}, \citenamefont {Ruddock},
  \citenamefont {Shen}, \citenamefont {Vecchione}, \citenamefont {Weathersby},
  \citenamefont {Weber}, \citenamefont {Wilkin}, \citenamefont {Yong},
  \citenamefont {Zheng}, \citenamefont {Wang}, \citenamefont {Minitti}, and\
  \citenamefont {Martinez}}]{Wolf:NatChem11:504}%
  \BibitemOpen
  \bibfield  {author} {\bibinfo {author} {\bibfnamefont {T.~J.~A.}\
  \bibnamefont {Wolf}}, \bibinfo {author} {\bibfnamefont {D.~M.}\ \bibnamefont
  {Sanchez}}, \bibinfo {author} {\bibfnamefont {J.}~\bibnamefont {Yang}},
  \bibinfo {author} {\bibfnamefont {R.~M.}\ \bibnamefont {Parrish}}, \bibinfo
  {author} {\bibfnamefont {J.~P.~F.}\ \bibnamefont {Nunes}}, \bibinfo {author}
  {\bibfnamefont {M.}~\bibnamefont {Centurion}}, \bibinfo {author}
  {\bibfnamefont {R.}~\bibnamefont {Coffee}}, \bibinfo {author} {\bibfnamefont
  {J.~P.}\ \bibnamefont {Cryan}}, \bibinfo {author} {\bibfnamefont
  {M.}~\bibnamefont {G{\"u}hr}}, \bibinfo {author} {\bibfnamefont
  {K.}~\bibnamefont {Hegazy}}, \bibinfo {author} {\bibfnamefont
  {A.}~\bibnamefont {Kirrander}}, \bibinfo {author} {\bibfnamefont {R.~K.}\
  \bibnamefont {Li}}, \bibinfo {author} {\bibfnamefont {J.}~\bibnamefont
  {Ruddock}}, \bibinfo {author} {\bibfnamefont {X.}~\bibnamefont {Shen}},
  \bibinfo {author} {\bibfnamefont {T.}~\bibnamefont {Vecchione}}, \bibinfo
  {author} {\bibfnamefont {S.~P.}\ \bibnamefont {Weathersby}}, \bibinfo
  {author} {\bibfnamefont {P.~M.}\ \bibnamefont {Weber}}, \bibinfo {author}
  {\bibfnamefont {K.}~\bibnamefont {Wilkin}}, \bibinfo {author} {\bibfnamefont
  {H.}~\bibnamefont {Yong}}, \bibinfo {author} {\bibfnamefont {Q.}~\bibnamefont
  {Zheng}}, \bibinfo {author} {\bibfnamefont {X.~J.}\ \bibnamefont {Wang}},
  \bibinfo {author} {\bibfnamefont {M.~P.}\ \bibnamefont {Minitti}}, and\
  \bibinfo {author} {\bibfnamefont {T.~J.}\ \bibnamefont {Martinez}},\
  }\bibfield  {title} {\enquote {\bibinfo {title} {The photochemical
  ring-opening of 1,3-cyclohexadiene imaged by ultrafast electron
  diffraction},}\ }\href {https://doi.org/10.1038/s41557-019-0252-7} {\bibfield
   {journal} {\bibinfo  {journal} {Nat. Chem.}\ }\textbf {\bibinfo {volume}
  {11}},\ \bibinfo {pages} {504--509} (\bibinfo {year} {2019})}\BibitemShut
  {NoStop}%
\bibitem [{\citenamefont {Karamatskos}\ \emph
  {et~al.}(2019{\natexlab{a}})\citenamefont {Karamatskos}, \citenamefont
  {Goldsztejn}, \citenamefont {Raabe}, \citenamefont {Stammer}, \citenamefont
  {Mullins}, \citenamefont {Trabattoni}, \citenamefont {Johansen},
  \citenamefont {Stapelfeldt}, \citenamefont {Trippel}, \citenamefont
  {Vrakking}, \citenamefont {K{\"u}pper}, and\ \citenamefont
  {Rouz{\'e}e}}]{Karamatskos:JCP150:244301}%
  \BibitemOpen
  \bibfield  {author} {\bibinfo {author} {\bibfnamefont {E.~T.}\ \bibnamefont
  {Karamatskos}}, \bibinfo {author} {\bibfnamefont {G.}~\bibnamefont
  {Goldsztejn}}, \bibinfo {author} {\bibfnamefont {S.}~\bibnamefont {Raabe}},
  \bibinfo {author} {\bibfnamefont {P.}~\bibnamefont {Stammer}}, \bibinfo
  {author} {\bibfnamefont {T.}~\bibnamefont {Mullins}}, \bibinfo {author}
  {\bibfnamefont {A.}~\bibnamefont {Trabattoni}}, \bibinfo {author}
  {\bibfnamefont {R.~R.}\ \bibnamefont {Johansen}}, \bibinfo {author}
  {\bibfnamefont {H.}~\bibnamefont {Stapelfeldt}}, \bibinfo {author}
  {\bibfnamefont {S.}~\bibnamefont {Trippel}}, \bibinfo {author} {\bibfnamefont
  {M.~J.~J.}\ \bibnamefont {Vrakking}}, \bibinfo {author} {\bibfnamefont
  {J.}~\bibnamefont {K{\"u}pper}}, and\ \bibinfo {author} {\bibfnamefont
  {A.}~\bibnamefont {Rouz{\'e}e}},\ }\bibfield  {title} {\enquote {\bibinfo
  {title} {Atomic-resolution imaging of carbonyl sulfide by laser-induced
  electron diffraction},}\ }\href {https://doi.org/10.1063/1.5093959}
  {\bibfield  {journal} {\bibinfo  {journal} {J. Chem. Phys.}\ }\textbf
  {\bibinfo {volume} {150}},\ \bibinfo {pages} {244301} (\bibinfo {year}
  {2019}{\natexlab{a}})},\ \Eprint {https://arxiv.org/abs/1905.03541}
  {arXiv:1905.03541 [physics]}\BibitemShut {NoStop}%
\bibitem [{\citenamefont {Trabattoni}\ \emph {et~al.}(2020)\citenamefont
  {Trabattoni}, \citenamefont {Wiese}, \citenamefont {De~Giovannini},
  \citenamefont {Olivieri}, \citenamefont {Mullins}, \citenamefont {Onvlee},
  \citenamefont {Son}, \citenamefont {Frusteri}, \citenamefont {Rubio},
  \citenamefont {Trippel}, and\ \citenamefont
  {Küpper}}]{Trabattoni:NatComm11:2546}%
  \BibitemOpen
  \bibfield  {author} {\bibinfo {author} {\bibfnamefont {A.}~\bibnamefont
  {Trabattoni}}, \bibinfo {author} {\bibfnamefont {J.}~\bibnamefont {Wiese}},
  \bibinfo {author} {\bibfnamefont {U.}~\bibnamefont {De~Giovannini}}, \bibinfo
  {author} {\bibfnamefont {J.-F.}\ \bibnamefont {Olivieri}}, \bibinfo {author}
  {\bibfnamefont {T.}~\bibnamefont {Mullins}}, \bibinfo {author} {\bibfnamefont
  {J.}~\bibnamefont {Onvlee}}, \bibinfo {author} {\bibfnamefont {S.-K.}\
  \bibnamefont {Son}}, \bibinfo {author} {\bibfnamefont {B.}~\bibnamefont
  {Frusteri}}, \bibinfo {author} {\bibfnamefont {A.}~\bibnamefont {Rubio}},
  \bibinfo {author} {\bibfnamefont {S.}~\bibnamefont {Trippel}}, and\ \bibinfo
  {author} {\bibfnamefont {J.}~\bibnamefont {Küpper}},\ }\bibfield  {title}
  {\enquote {\bibinfo {title} {Setting the photoelectron clock through
  molecular alignment},}\ }\href {https://doi.org/10.1038/s41467-020-16270-0}
  {\bibfield  {journal} {\bibinfo  {journal} {Nat. Commun.}\ }\textbf {\bibinfo
  {volume} {11}},\ \bibinfo {pages} {2546} (\bibinfo {year} {2020})},\ \Eprint
  {https://arxiv.org/abs/1802.06622} {arXiv:1802.06622 [physics]}\BibitemShut
  {NoStop}%
\bibitem [{\citenamefont {Karamatskos}(2019)}]{Karamatskos:thesis:2019}%
  \BibitemOpen
  \bibfield  {author} {\bibinfo {author} {\bibfnamefont {E.}~\bibnamefont
  {Karamatskos}},\ }\emph {\bibinfo {title} {Molecular-Frame Angularly-Resolved
  Photoelectron Spectroscopy}},\ \href
  {https://ediss.sub.uni-hamburg.de/volltexte/2019/10050/} {\bibinfo {type}
  {Dissertation}},\ \bibinfo  {school} {Universit{\"a}t Hamburg}, \bibinfo
  {address} {{H}amburg, Germany} (\bibinfo {year} {2019})\BibitemShut {NoStop}%
\bibitem [{\citenamefont {Rabalais}\ \emph {et~al.}(1971)\citenamefont
  {Rabalais}, \citenamefont {McDonald}, \citenamefont {Scherr}, and\
  \citenamefont {McGlynn}}]{Rabalais:CR71:73}%
  \BibitemOpen
  \bibfield  {author} {\bibinfo {author} {\bibfnamefont {J.~W.}\ \bibnamefont
  {Rabalais}}, \bibinfo {author} {\bibfnamefont {J.~M.}\ \bibnamefont
  {McDonald}}, \bibinfo {author} {\bibfnamefont {V.}~\bibnamefont {Scherr}},\
  and\ \bibinfo {author} {\bibfnamefont {S.~P.}\ \bibnamefont {McGlynn}},\
  }\bibfield  {title} {\enquote {\bibinfo {title} {Electronic spectroscopy of
  isoelectronic molecules. {II}. {L}inear triatomic groupings containing
  sixteen valence electrons},}\ }\href {https://doi.org/10.1021/cr60269a004}
  {\bibfield  {journal} {\bibinfo  {journal} {Chem. Rev.}\ }\textbf {\bibinfo
  {volume} {71}},\ \bibinfo {pages} {73--108} (\bibinfo {year}
  {1971})}\BibitemShut {NoStop}%
\bibitem [{\citenamefont {Sivakumar}\ \emph {et~al.}(1988)\citenamefont
  {Sivakumar}, \citenamefont {Hall}, \citenamefont {Houston}, \citenamefont
  {Hepburn}, and\ \citenamefont {Burak}}]{Sivakumar:JCP88:3692}%
  \BibitemOpen
  \bibfield  {author} {\bibinfo {author} {\bibfnamefont {N.}~\bibnamefont
  {Sivakumar}}, \bibinfo {author} {\bibfnamefont {G.~E.}\ \bibnamefont {Hall}},
  \bibinfo {author} {\bibfnamefont {P.~L.}\ \bibnamefont {Houston}}, \bibinfo
  {author} {\bibfnamefont {J.~W.}\ \bibnamefont {Hepburn}}, and\ \bibinfo
  {author} {\bibfnamefont {I.}~\bibnamefont {Burak}},\ }\bibfield  {title}
  {\enquote {\bibinfo {title} {State-resolved photodissociation of {OCS}
  monomers and clusters},}\ }\href {https://doi.org/10.1063/1.453869}
  {\bibfield  {journal} {\bibinfo  {journal} {J. Chem. Phys.}\ }\textbf
  {\bibinfo {volume} {88}},\ \bibinfo {pages} {3692--3708} (\bibinfo {year}
  {1988})}\BibitemShut {NoStop}%
\bibitem [{\citenamefont {Sato}\ \emph {et~al.}(1995)\citenamefont {Sato},
  \citenamefont {Matsumi}, \citenamefont {Kawasaki}, \citenamefont
  {Tsukiyama}, and\ \citenamefont {Bersohn}}]{Sato:JPC99:16307}%
  \BibitemOpen
  \bibfield  {author} {\bibinfo {author} {\bibfnamefont {Y.}~\bibnamefont
  {Sato}}, \bibinfo {author} {\bibfnamefont {Y.}~\bibnamefont {Matsumi}},
  \bibinfo {author} {\bibfnamefont {M.}~\bibnamefont {Kawasaki}}, \bibinfo
  {author} {\bibfnamefont {K.}~\bibnamefont {Tsukiyama}}, and\ \bibinfo
  {author} {\bibfnamefont {R.}~\bibnamefont {Bersohn}},\ }\bibfield  {title}
  {\enquote {\bibinfo {title} {Ion imaging of the photodissociation of {OCS}
  near 217 and 230~nm},}\ }\href {https://doi.org/10.1021/j100044a017}
  {\bibfield  {journal} {\bibinfo  {journal} {J. Phys. Chem.}\ }\textbf
  {\bibinfo {volume} {99}},\ \bibinfo {pages} {16307--16314} (\bibinfo {year}
  {1995})}\BibitemShut {NoStop}%
\bibitem [{\citenamefont {Suzuki}\ \emph {et~al.}(1998)\citenamefont {Suzuki},
  \citenamefont {Katayanagi}, \citenamefont {Nanbu}, and\ \citenamefont
  {Aoyagi}}]{Suzuki:JCP109:5778}%
  \BibitemOpen
  \bibfield  {author} {\bibinfo {author} {\bibfnamefont {T.}~\bibnamefont
  {Suzuki}}, \bibinfo {author} {\bibfnamefont {H.}~\bibnamefont {Katayanagi}},
  \bibinfo {author} {\bibfnamefont {S.}~\bibnamefont {Nanbu}}, and\ \bibinfo
  {author} {\bibfnamefont {M.}~\bibnamefont {Aoyagi}},\ }\bibfield  {title}
  {\enquote {\bibinfo {title} {Nonadiabatic bending dissociation in 16 valence
  electron system {OCS}},}\ }\href {https://doi.org/10.1063/1.477200}
  {\bibfield  {journal} {\bibinfo  {journal} {J. Chem. Phys.}\ }\textbf
  {\bibinfo {volume} {109}},\ \bibinfo {pages} {5778--5794} (\bibinfo {year}
  {1998})}\BibitemShut {NoStop}%
\bibitem [{\citenamefont {Breckenridge} and\ \citenamefont
  {Taube}(1970)}]{Breckenridge:JCP52:1713}%
  \BibitemOpen
  \bibfield  {author} {\bibinfo {author} {\bibfnamefont {W.~H.}\ \bibnamefont
  {Breckenridge}} and\ \bibinfo {author} {\bibfnamefont {H.}~\bibnamefont
  {Taube}},\ }\bibfield  {title} {\enquote {\bibinfo {title} {Ultraviolet
  absorption spectrum of carbonyl sulfide},}\ }\href
  {https://doi.org/10.1063/1.1673209} {\bibfield  {journal} {\bibinfo
  {journal} {J. Chem. Phys.}\ }\textbf {\bibinfo {volume} {52}},\ \bibinfo
  {pages} {1713--1715} (\bibinfo {year} {1970})}\BibitemShut {NoStop}%
\bibitem [{\citenamefont {Sivakumar}\ \emph {et~al.}(1985)\citenamefont
  {Sivakumar}, \citenamefont {Burak}, \citenamefont {Cheung}, \citenamefont
  {Houston}, and\ \citenamefont {Hepburn}}]{Sivakumar:JPC89:3609}%
  \BibitemOpen
  \bibfield  {author} {\bibinfo {author} {\bibfnamefont {N.}~\bibnamefont
  {Sivakumar}}, \bibinfo {author} {\bibfnamefont {I.}~\bibnamefont {Burak}},
  \bibinfo {author} {\bibfnamefont {W.~Y.}\ \bibnamefont {Cheung}}, \bibinfo
  {author} {\bibfnamefont {P.~L.}\ \bibnamefont {Houston}}, and\ \bibinfo
  {author} {\bibfnamefont {J.~W.}\ \bibnamefont {Hepburn}},\ }\bibfield
  {title} {\enquote {\bibinfo {title} {State-resolved photofragmentation of
  carbonyl sulfide {(OCS)} monomers and clusters},}\ }\href
  {https://doi.org/10.1021/j100263a008} {\bibfield  {journal} {\bibinfo
  {journal} {J. Phys. Chem.}\ }\textbf {\bibinfo {volume} {89}},\ \bibinfo
  {pages} {3609--3611} (\bibinfo {year} {1985})}\BibitemShut {NoStop}%
\bibitem [{\citenamefont {Katayanagi}, \citenamefont {Mo}, and\ \citenamefont
  {Suzuki}(1995)}]{Katayanagi:CPL247:571}%
  \BibitemOpen
  \bibfield  {author} {\bibinfo {author} {\bibfnamefont {H.}~\bibnamefont
  {Katayanagi}}, \bibinfo {author} {\bibfnamefont {Y.}~\bibnamefont {Mo}},\
  and\ \bibinfo {author} {\bibfnamefont {T.}~\bibnamefont {Suzuki}},\
  }\bibfield  {title} {\enquote {\bibinfo {title} {223~nm photodissociation of
  {OCS}. {T}wo components in {S($^1\!D_2$)} and {S($^3\!P_2$)} channels},}\
  }\href {https://doi.org/https://doi.org/10.1016/S0009-2614(95)01253-2}
  {\bibfield  {journal} {\bibinfo  {journal} {Chem. Phys. Lett.}\ }\textbf
  {\bibinfo {volume} {247}},\ \bibinfo {pages} {571--576} (\bibinfo {year}
  {1995})}\BibitemShut {NoStop}%
\bibitem [{\citenamefont {Kim}, \citenamefont {Alexander}, and\ \citenamefont
  {Zare}(1999)}]{Kim:JPCA103:10144}%
  \BibitemOpen
  \bibfield  {author} {\bibinfo {author} {\bibfnamefont {Z.~H.}\ \bibnamefont
  {Kim}}, \bibinfo {author} {\bibfnamefont {A.~J.}\ \bibnamefont {Alexander}},\
  and\ \bibinfo {author} {\bibfnamefont {R.~N.}\ \bibnamefont {Zare}},\
  }\bibfield  {title} {\enquote {\bibinfo {title} {Speed-dependent
  photofragment orientation in the photodissociation of {OCS} at 223~nm},}\
  }\href {https://doi.org/10.1021/jp991988q} {\bibfield  {journal} {\bibinfo
  {journal} {J. Phys. Chem. A}\ }\textbf {\bibinfo {volume} {103}},\ \bibinfo
  {pages} {10144--10148} (\bibinfo {year} {1999})}\BibitemShut {NoStop}%
\bibitem [{\citenamefont {Sugita}\ \emph {et~al.}(2000)\citenamefont {Sugita},
  \citenamefont {Mashino}, \citenamefont {Kawasaki}, \citenamefont {Matsumi},
  \citenamefont {Bersohn}, \citenamefont {Trott-Kriegeskorte}, and\
  \citenamefont {Gericke}}]{Sugita:JCP112:7095}%
  \BibitemOpen
  \bibfield  {author} {\bibinfo {author} {\bibfnamefont {A.}~\bibnamefont
  {Sugita}}, \bibinfo {author} {\bibfnamefont {M.}~\bibnamefont {Mashino}},
  \bibinfo {author} {\bibfnamefont {M.}~\bibnamefont {Kawasaki}}, \bibinfo
  {author} {\bibfnamefont {Y.}~\bibnamefont {Matsumi}}, \bibinfo {author}
  {\bibfnamefont {R.}~\bibnamefont {Bersohn}}, \bibinfo {author} {\bibfnamefont
  {G.}~\bibnamefont {Trott-Kriegeskorte}}, and\ \bibinfo {author}
  {\bibfnamefont {K.~a.-H.}\ \bibnamefont {Gericke}},\ }\bibfield  {title}
  {\enquote {\bibinfo {title} {Effect of molecular bending on the
  photodissociation of {OCS}},}\ }\href {https://doi.org/10.1063/1.481324}
  {\bibfield  {journal} {\bibinfo  {journal} {J. Chem. Phys.}\ }\textbf
  {\bibinfo {volume} {112}},\ \bibinfo {pages} {7095--7101} (\bibinfo {year}
  {2000})}\BibitemShut {NoStop}%
\bibitem [{\citenamefont {Rakitzis}, \citenamefont {Samartzis}, and\
  \citenamefont {Kitsopoulos}(1999)}]{Rakitzis:JCP111:10415}%
  \BibitemOpen
  \bibfield  {author} {\bibinfo {author} {\bibfnamefont {T.~P.}\ \bibnamefont
  {Rakitzis}}, \bibinfo {author} {\bibfnamefont {P.~C.}\ \bibnamefont
  {Samartzis}}, and\ \bibinfo {author} {\bibfnamefont {T.~N.}\ \bibnamefont
  {Kitsopoulos}},\ }\bibfield  {title} {\enquote {\bibinfo {title} {Observing
  the symmetry breaking in the angular distributions of oriented photofragments
  using velocity mapping},}\ }\href {https://doi.org/10.1063/1.480396}
  {\bibfield  {journal} {\bibinfo  {journal} {J. Chem. Phys.}\ }\textbf
  {\bibinfo {volume} {111}},\ \bibinfo {pages} {10415--10417} (\bibinfo {year}
  {1999})}\BibitemShut {NoStop}%
\bibitem [{\citenamefont {Katayanagi} and\ \citenamefont
  {Suzuki}(2002)}]{Katayanagi:CPL360:104}%
  \BibitemOpen
  \bibfield  {author} {\bibinfo {author} {\bibfnamefont {H.}~\bibnamefont
  {Katayanagi}} and\ \bibinfo {author} {\bibfnamefont {T.}~\bibnamefont
  {Suzuki}},\ }\bibfield  {title} {\enquote {\bibinfo {title} {Non-adiabatic
  bending dissociation of {OCS}: the effect of bending excitation on the
  transition probability},}\ }\href
  {https://doi.org/10.1016/S0009-2614(02)00788-1} {\bibfield  {journal}
  {\bibinfo  {journal} {Chem. Phys. Lett.}\ }\textbf {\bibinfo {volume}
  {360}},\ \bibinfo {pages} {104--110} (\bibinfo {year} {2002})}\BibitemShut
  {NoStop}%
\bibitem [{\citenamefont {Rijs}\ \emph {et~al.}(2002)\citenamefont {Rijs},
  \citenamefont {Backus}, \citenamefont {de~Lange}, \citenamefont {Janssen},
  \citenamefont {Westwood}, \citenamefont {Wang}, and\ \citenamefont
  {McKoy}}]{Rijs:JCP116:2776}%
  \BibitemOpen
  \bibfield  {author} {\bibinfo {author} {\bibfnamefont {A.~M.}\ \bibnamefont
  {Rijs}}, \bibinfo {author} {\bibfnamefont {E.~H.~G.}\ \bibnamefont {Backus}},
  \bibinfo {author} {\bibfnamefont {C.~A.}\ \bibnamefont {de~Lange}}, \bibinfo
  {author} {\bibfnamefont {M.~H.~M.}\ \bibnamefont {Janssen}}, \bibinfo
  {author} {\bibfnamefont {N.~P.~C.}\ \bibnamefont {Westwood}}, \bibinfo
  {author} {\bibfnamefont {K.}~\bibnamefont {Wang}}, and\ \bibinfo {author}
  {\bibfnamefont {V.}~\bibnamefont {McKoy}},\ }\bibfield  {title} {\enquote
  {\bibinfo {title} {Rotationally resolved photoionization dynamics of hot {CO}
  fragmented from {OCS}},}\ }\href {https://doi.org/10.1063/1.1434993}
  {\bibfield  {journal} {\bibinfo  {journal} {J. Chem. Phys.}\ }\textbf
  {\bibinfo {volume} {116}},\ \bibinfo {pages} {2776--2782} (\bibinfo {year}
  {2002})}\BibitemShut {NoStop}%
\bibitem [{\citenamefont {Wei}\ \emph {et~al.}(2016)\citenamefont {Wei},
  \citenamefont {Wallace}, \citenamefont {McBane}, and\ \citenamefont
  {North}}]{Wei:JCP145:024310}%
  \BibitemOpen
  \bibfield  {author} {\bibinfo {author} {\bibfnamefont {W.}~\bibnamefont
  {Wei}}, \bibinfo {author} {\bibfnamefont {C.~J.}\ \bibnamefont {Wallace}},
  \bibinfo {author} {\bibfnamefont {G.~C.}\ \bibnamefont {McBane}}, and\
  \bibinfo {author} {\bibfnamefont {S.~W.}\ \bibnamefont {North}},\ }\bibfield
  {title} {\enquote {\bibinfo {title} {Photodissociation dynamics of {OCS} near
  214 nm using ion imaging},}\ }\href {https://doi.org/10.1063/1.4955189}
  {\bibfield  {journal} {\bibinfo  {journal} {J. Chem. Phys.}\ }\textbf
  {\bibinfo {volume} {145}},\ \bibinfo {pages} {024310} (\bibinfo {year}
  {2016})}\BibitemShut {NoStop}%
\bibitem [{\citenamefont {Brouard}\ \emph {et~al.}(2007)\citenamefont
  {Brouard}, \citenamefont {Green}, \citenamefont {Quadrini}, and\
  \citenamefont {Vallance}}]{Brouard:JCP127:084304}%
  \BibitemOpen
  \bibfield  {author} {\bibinfo {author} {\bibfnamefont {M.}~\bibnamefont
  {Brouard}}, \bibinfo {author} {\bibfnamefont {A.~V.}\ \bibnamefont {Green}},
  \bibinfo {author} {\bibfnamefont {F.}~\bibnamefont {Quadrini}}, and\
  \bibinfo {author} {\bibfnamefont {C.}~\bibnamefont {Vallance}},\ }\bibfield
  {title} {\enquote {\bibinfo {title} {Photodissociation dynamics of {OCS} at
  248~nm: The {S($^1\!D_2$)} atomic angular momentum polarization},}\ }\href
  {https://doi.org/10.1063/1.2757618} {\bibfield  {journal} {\bibinfo
  {journal} {J. Chem. Phys.}\ }\textbf {\bibinfo {volume} {127}},\ \bibinfo
  {pages} {084304} (\bibinfo {year} {2007})}\BibitemShut {NoStop}%
\bibitem [{\citenamefont {van~den Brom}, \citenamefont {Rakitzis}, and\
  \citenamefont {Janssen}(2004)}]{VanDenBrom:JCP121:11645}%
  \BibitemOpen
  \bibfield  {author} {\bibinfo {author} {\bibfnamefont {A.~J.}\ \bibnamefont
  {van~den Brom}}, \bibinfo {author} {\bibfnamefont {T.~P.}\ \bibnamefont
  {Rakitzis}}, and\ \bibinfo {author} {\bibfnamefont {M.~H.~M.}\ \bibnamefont
  {Janssen}},\ }\bibfield  {title} {\enquote {\bibinfo {title}
  {Photodissociation of laboratory oriented molecules: {R}evealing molecular
  frame properties of nonaxial recoil},}\ }\href
  {https://doi.org/10.1063/1.1812756} {\bibfield  {journal} {\bibinfo
  {journal} {J. Chem. Phys.}\ }\textbf {\bibinfo {volume} {121}},\ \bibinfo
  {pages} {11645--11652} (\bibinfo {year} {2004})}\BibitemShut {NoStop}%
\bibitem [{\citenamefont {van~den Brom}, \citenamefont {Rakitzis}, and\
  \citenamefont {Janssen}(2005)}]{VanDenBrom:JCP123:164313}%
  \BibitemOpen
  \bibfield  {author} {\bibinfo {author} {\bibfnamefont {A.~J.}\ \bibnamefont
  {van~den Brom}}, \bibinfo {author} {\bibfnamefont {T.~P.}\ \bibnamefont
  {Rakitzis}}, and\ \bibinfo {author} {\bibfnamefont {M.~H.~M.}\ \bibnamefont
  {Janssen}},\ }\bibfield  {title} {\enquote {\bibinfo {title} {State-to-state
  photodissociation of carbonyl sulfide $(v_2=0,1|{JIM})$. {II}. {T}he effect
  of initial bending on coherence of {S($^1\!D_2$)} polarization},}\ }\href
  {https://doi.org/10.1063/1.2076647} {\bibfield  {journal} {\bibinfo
  {journal} {J. Chem. Phys.}\ }\textbf {\bibinfo {volume} {123}},\ \bibinfo
  {pages} {164313} (\bibinfo {year} {2005})}\BibitemShut {NoStop}%
\bibitem [{\citenamefont {Rakitzis}, \citenamefont {van~den Brom}, and\
  \citenamefont {Janssen}(2004)}]{Rakitzis:Science303:1852}%
  \BibitemOpen
  \bibfield  {author} {\bibinfo {author} {\bibfnamefont {T.~P.}\ \bibnamefont
  {Rakitzis}}, \bibinfo {author} {\bibfnamefont {A.~J.}\ \bibnamefont {van~den
  Brom}}, and\ \bibinfo {author} {\bibfnamefont {M.~H.~M.}\ \bibnamefont
  {Janssen}},\ }\bibfield  {title} {\enquote {\bibinfo {title} {Directional
  dynamics in the photodissociation of oriented molecules},}\ }\href
  {https://doi.org/10.1126/science.1094186} {\bibfield  {journal} {\bibinfo
  {journal} {Science}\ }\textbf {\bibinfo {volume} {303}},\ \bibinfo {pages}
  {1852--1854} (\bibinfo {year} {2004})}\BibitemShut {NoStop}%
\bibitem [{\citenamefont {Lipciuc} and\ \citenamefont
  {Janssen}(2006)}]{Lipciuc:PCCP8:3007}%
  \BibitemOpen
  \bibfield  {author} {\bibinfo {author} {\bibfnamefont {M.~L.}\ \bibnamefont
  {Lipciuc}} and\ \bibinfo {author} {\bibfnamefont {M.~H.~M.}\ \bibnamefont
  {Janssen}},\ }\bibfield  {title} {\enquote {\bibinfo {title} {Slice imaging
  of quantum state-to-state photodissociation dynamics of {OCS}},}\ }\href
  {https://doi.org/10.1039/B605108A} {\bibfield  {journal} {\bibinfo  {journal}
  {Phys. Chem. Chem. Phys.}\ }\textbf {\bibinfo {volume} {8}},\ \bibinfo
  {pages} {3007--3016} (\bibinfo {year} {2006})}\BibitemShut {NoStop}%
\bibitem [{\citenamefont {Lipciuc}\ \emph {et~al.}(2011)\citenamefont
  {Lipciuc}, \citenamefont {Rakitzis}, \citenamefont {Meerts}, \citenamefont
  {Groenenboom}, and\ \citenamefont {Janssen}}]{Lipciuc:PCCP13:8549}%
  \BibitemOpen
  \bibfield  {author} {\bibinfo {author} {\bibfnamefont {M.~L.}\ \bibnamefont
  {Lipciuc}}, \bibinfo {author} {\bibfnamefont {T.~P.}\ \bibnamefont
  {Rakitzis}}, \bibinfo {author} {\bibfnamefont {W.~L.}\ \bibnamefont
  {Meerts}}, \bibinfo {author} {\bibfnamefont {G.~C.}\ \bibnamefont
  {Groenenboom}}, and\ \bibinfo {author} {\bibfnamefont {M.~H.~M.}\
  \bibnamefont {Janssen}},\ }\bibfield  {title} {\enquote {\bibinfo {title}
  {Towards the complete experiment: measurement of {S($^1\!D_2$)} polarization
  in correlation with single rotational states of {CO$(J)$} from the
  photodissociation of oriented {OCS}$(v_2=1|{JlM}=111)$},}\ }\href
  {https://doi.org/10.1039/C0CP02671A} {\bibfield  {journal} {\bibinfo
  {journal} {Phys. Chem. Chem. Phys.}\ }\textbf {\bibinfo {volume} {13}},\
  \bibinfo {pages} {8549--8559} (\bibinfo {year} {2011})}\BibitemShut {NoStop}%
\bibitem [{\citenamefont {Sofikitis}\ \emph {et~al.}(2017)\citenamefont
  {Sofikitis}, \citenamefont {Suarez}, \citenamefont {Schmidt}, \citenamefont
  {Rakitzis}, \citenamefont {Farantos}, and\ \citenamefont
  {Janssen}}]{Sofikitis:PRL118:253001}%
  \BibitemOpen
  \bibfield  {author} {\bibinfo {author} {\bibfnamefont {D.}~\bibnamefont
  {Sofikitis}}, \bibinfo {author} {\bibfnamefont {J.}~\bibnamefont {Suarez}},
  \bibinfo {author} {\bibfnamefont {J.~A.}\ \bibnamefont {Schmidt}}, \bibinfo
  {author} {\bibfnamefont {T.~P.}\ \bibnamefont {Rakitzis}}, \bibinfo {author}
  {\bibfnamefont {S.~C.}\ \bibnamefont {Farantos}}, and\ \bibinfo {author}
  {\bibfnamefont {M.~H.~M.}\ \bibnamefont {Janssen}},\ }\bibfield  {title}
  {\enquote {\bibinfo {title} {Recoil inversion in the photodissociation of
  carbonyl sulfide near 234 nm},}\ }\href
  {https://doi.org/10.1103/PhysRevLett.118.253001} {\bibfield  {journal}
  {\bibinfo  {journal} {Phys. Rev. Lett.}\ }\textbf {\bibinfo {volume} {118}},\
  \bibinfo {pages} {253001} (\bibinfo {year} {2017})}\BibitemShut {NoStop}%
\bibitem [{\citenamefont {Sofikitis}\ \emph {et~al.}(2018)\citenamefont
  {Sofikitis}, \citenamefont {Suarez}, \citenamefont {Schmidt}, \citenamefont
  {Rakitzis}, \citenamefont {Farantos}, and\ \citenamefont
  {Janssen}}]{Sofikitis:PRA98:033417}%
  \BibitemOpen
  \bibfield  {author} {\bibinfo {author} {\bibfnamefont {D.}~\bibnamefont
  {Sofikitis}}, \bibinfo {author} {\bibfnamefont {J.}~\bibnamefont {Suarez}},
  \bibinfo {author} {\bibfnamefont {J.~A.}\ \bibnamefont {Schmidt}}, \bibinfo
  {author} {\bibfnamefont {T.~P.}\ \bibnamefont {Rakitzis}}, \bibinfo {author}
  {\bibfnamefont {S.~C.}\ \bibnamefont {Farantos}}, and\ \bibinfo {author}
  {\bibfnamefont {M.~H.~M.}\ \bibnamefont {Janssen}},\ }\bibfield  {title}
  {\enquote {\bibinfo {title} {Exit-channel recoil resonances by imaging the
  photodissociation of single quantum-state-selected {OCS} molecules},}\ }\href
  {https://doi.org/10.1103/PhysRevA.98.033417} {\bibfield  {journal} {\bibinfo
  {journal} {Phys. Rev. A}\ }\textbf {\bibinfo {volume} {98}},\ \bibinfo
  {pages} {033417} (\bibinfo {year} {2018})}\BibitemShut {NoStop}%
\bibitem [{\citenamefont {Schmidt}\ \emph
  {et~al.}(2012{\natexlab{a}})\citenamefont {Schmidt}, \citenamefont {Johnson},
  \citenamefont {McBane}, and\ \citenamefont
  {Schinke}}]{Schmidt:JCP137:054313}%
  \BibitemOpen
  \bibfield  {author} {\bibinfo {author} {\bibfnamefont {J.~A.}\ \bibnamefont
  {Schmidt}}, \bibinfo {author} {\bibfnamefont {M.~S.}\ \bibnamefont
  {Johnson}}, \bibinfo {author} {\bibfnamefont {G.~C.}\ \bibnamefont
  {McBane}}, and\ \bibinfo {author} {\bibfnamefont {R.}~\bibnamefont
  {Schinke}},\ }\bibfield  {title} {\enquote {\bibinfo {title} {The ultraviolet
  spectrum of {OCS} from first principles: Electronic transitions, vibrational
  structure and temperature dependence},}\ }\href
  {https://doi.org/10.1063/1.4739756} {\bibfield  {journal} {\bibinfo
  {journal} {J. Chem. Phys.}\ }\textbf {\bibinfo {volume} {137}},\ \bibinfo
  {pages} {054313} (\bibinfo {year} {2012}{\natexlab{a}})}\BibitemShut
  {NoStop}%
\bibitem [{\citenamefont {Schmidt}\ \emph {et~al.}(2013)\citenamefont
  {Schmidt}, \citenamefont {Johnson}, \citenamefont {Hattori}, \citenamefont
  {Yoshida}, \citenamefont {Nanbu}, and\ \citenamefont
  {Schinke}}]{Schmidt:ACP13:1511}%
  \BibitemOpen
  \bibfield  {author} {\bibinfo {author} {\bibfnamefont {J.~A.}\ \bibnamefont
  {Schmidt}}, \bibinfo {author} {\bibfnamefont {M.~S.}\ \bibnamefont
  {Johnson}}, \bibinfo {author} {\bibfnamefont {S.}~\bibnamefont {Hattori}},
  \bibinfo {author} {\bibfnamefont {N.}~\bibnamefont {Yoshida}}, \bibinfo
  {author} {\bibfnamefont {S.}~\bibnamefont {Nanbu}}, and\ \bibinfo {author}
  {\bibfnamefont {R.}~\bibnamefont {Schinke}},\ }\bibfield  {title} {\enquote
  {\bibinfo {title} {{OCS} photolytic isotope effects from first principles:
  sulfur and carbon isotopes, temperature dependence and implications for the
  stratosphere},}\ }\href {https://doi.org/10.5194/acp-13-1511-2013} {\bibfield
   {journal} {\bibinfo  {journal} {Atmos. Chem. Phys.}\ }\textbf {\bibinfo
  {volume} {13}},\ \bibinfo {pages} {1511--1520} (\bibinfo {year}
  {2013})}\BibitemShut {NoStop}%
\bibitem [{\citenamefont {Schmidt}\ \emph
  {et~al.}(2012{\natexlab{b}})\citenamefont {Schmidt}, \citenamefont {Johnson},
  \citenamefont {McBane}, and\ \citenamefont
  {Schinke}}]{Schmidt:JCP136:131101}%
  \BibitemOpen
  \bibfield  {author} {\bibinfo {author} {\bibfnamefont {J.~A.}\ \bibnamefont
  {Schmidt}}, \bibinfo {author} {\bibfnamefont {M.~S.}\ \bibnamefont
  {Johnson}}, \bibinfo {author} {\bibfnamefont {G.~C.}\ \bibnamefont
  {McBane}}, and\ \bibinfo {author} {\bibfnamefont {R.}~\bibnamefont
  {Schinke}},\ }\bibfield  {title} {\enquote {\bibinfo {title} {Multi-state
  analysis of the {OCS} ultraviolet absorption including vibrational
  structure},}\ }\href {https://doi.org/10.1063/1.3701699} {\bibfield
  {journal} {\bibinfo  {journal} {J. Chem. Phys.}\ }\textbf {\bibinfo {volume}
  {136}},\ \bibinfo {pages} {131101} (\bibinfo {year}
  {2012}{\natexlab{b}})}\BibitemShut {NoStop}%
\bibitem [{\citenamefont {McBane}\ \emph {et~al.}(2013)\citenamefont {McBane},
  \citenamefont {Schmidt}, \citenamefont {Johnson}, and\ \citenamefont
  {Schinke}}]{McBane:JCP138:094313}%
  \BibitemOpen
  \bibfield  {author} {\bibinfo {author} {\bibfnamefont {G.~C.}\ \bibnamefont
  {McBane}}, \bibinfo {author} {\bibfnamefont {J.~A.}\ \bibnamefont {Schmidt}},
  \bibinfo {author} {\bibfnamefont {M.~S.}\ \bibnamefont {Johnson}}, and\
  \bibinfo {author} {\bibfnamefont {R.}~\bibnamefont {Schinke}},\ }\bibfield
  {title} {\enquote {\bibinfo {title} {Ultraviolet photodissociation of {OCS}:
  Product energy and angular distributions},}\ }\href
  {https://doi.org/10.1063/1.4793275} {\bibfield  {journal} {\bibinfo
  {journal} {J. Chem. Phys.}\ }\textbf {\bibinfo {volume} {138}},\ \bibinfo
  {pages} {094314} (\bibinfo {year} {2013})}\BibitemShut {NoStop}%
\bibitem [{\citenamefont {Danielache}\ \emph {et~al.}(2009)\citenamefont
  {Danielache}, \citenamefont {Nanbu}, \citenamefont {Eskebjerg}, \citenamefont
  {Johnson}, and\ \citenamefont {Yoshida}}]{Danielache:JCP131:024307}%
  \BibitemOpen
  \bibfield  {author} {\bibinfo {author} {\bibfnamefont {S.~O.}\ \bibnamefont
  {Danielache}}, \bibinfo {author} {\bibfnamefont {S.}~\bibnamefont {Nanbu}},
  \bibinfo {author} {\bibfnamefont {C.}~\bibnamefont {Eskebjerg}}, \bibinfo
  {author} {\bibfnamefont {M.~S.}\ \bibnamefont {Johnson}}, and\ \bibinfo
  {author} {\bibfnamefont {N.}~\bibnamefont {Yoshida}},\ }\bibfield  {title}
  {\enquote {\bibinfo {title} {Carbonyl sulfide isotopologues: {U}ltraviolet
  absorption cross sections and stratospheric photolysis},}\ }\href
  {https://doi.org/10.1063/1.3156314} {\bibfield  {journal} {\bibinfo
  {journal} {J. Chem. Phys.}\ }\textbf {\bibinfo {volume} {131}},\ \bibinfo
  {pages} {024307} (\bibinfo {year} {2009})}\BibitemShut {NoStop}%
\bibitem [{\citenamefont {Brühl}\ \emph {et~al.}(2012)\citenamefont {Brühl},
  \citenamefont {Lelieveld}, \citenamefont {Crutzen}, and\ \citenamefont
  {Tost}}]{Buehl:AtmosChemPhys12:1239}%
  \BibitemOpen
  \bibfield  {author} {\bibinfo {author} {\bibfnamefont {C.}~\bibnamefont
  {Brühl}}, \bibinfo {author} {\bibfnamefont {J.}~\bibnamefont {Lelieveld}},
  \bibinfo {author} {\bibfnamefont {P.~J.}\ \bibnamefont {Crutzen}}, and\
  \bibinfo {author} {\bibfnamefont {H.}~\bibnamefont {Tost}},\ }\bibfield
  {title} {\enquote {\bibinfo {title} {The role of carbonyl sulphide as a
  source of stratospheric sulphate aerosol and its impact on climate},}\ }\href
  {https://doi.org/10.5194/acp-12-1239-2012} {\bibfield  {journal} {\bibinfo
  {journal} {Atmos. Chem. Phys.}\ }\textbf {\bibinfo {volume} {12}},\ \bibinfo
  {pages} {1239--1253} (\bibinfo {year} {2012})}\BibitemShut {NoStop}%
\bibitem [{\citenamefont {Hattori}, \citenamefont {Kamezaki}, and\
  \citenamefont {Yoshida}(2020)}]{Hattori:PNAS117:20447}%
  \BibitemOpen
  \bibfield  {author} {\bibinfo {author} {\bibfnamefont {S.}~\bibnamefont
  {Hattori}}, \bibinfo {author} {\bibfnamefont {K.}~\bibnamefont {Kamezaki}},\
  and\ \bibinfo {author} {\bibfnamefont {N.}~\bibnamefont {Yoshida}},\
  }\bibfield  {title} {\enquote {\bibinfo {title} {Constraining the atmospheric
  {OCS} budget from sulfur isotopes},}\ }\href
  {https://doi.org/10.1073/pnas.2007260117} {\bibfield  {journal} {\bibinfo
  {journal} {PNAS}\ }\textbf {\bibinfo {volume} {117}},\ \bibinfo {pages}
  {20447--20452} (\bibinfo {year} {2020})}\BibitemShut {NoStop}%
\bibitem [{\citenamefont {Lipciuc} and\ \citenamefont
  {Janssen}(2007)}]{Lipciuc:JCP126:194318}%
  \BibitemOpen
  \bibfield  {author} {\bibinfo {author} {\bibfnamefont {M.~L.}\ \bibnamefont
  {Lipciuc}} and\ \bibinfo {author} {\bibfnamefont {M.~H.~M.}\ \bibnamefont
  {Janssen}},\ }\bibfield  {title} {\enquote {\bibinfo {title} {Slice imaging
  of the quantum state-to-state cross section for photodissociation of
  state-selected rovibrational bending states of {OCS}
  {$(v_2=0,1,2|{JlM})+h\nu\rightarrow\text{CO}(J)+\text{S}(^1\!D_2)$}},}\
  }\href {https://doi.org/10.1063/1.2737450} {\bibfield  {journal} {\bibinfo
  {journal} {J. Chem. Phys.}\ }\textbf {\bibinfo {volume} {126}},\ \bibinfo
  {pages} {194318} (\bibinfo {year} {2007})}\BibitemShut {NoStop}%
\bibitem [{\citenamefont {Toulson} and\ \citenamefont
  {Murray}(2016)}]{Toulson:JPCA120:6745}%
  \BibitemOpen
  \bibfield  {author} {\bibinfo {author} {\bibfnamefont {B.~W.}\ \bibnamefont
  {Toulson}} and\ \bibinfo {author} {\bibfnamefont {C.}~\bibnamefont
  {Murray}},\ }\bibfield  {title} {\enquote {\bibinfo {title} {Decomposing the
  first absorption band of {OCS} using photofragment excitation
  spectroscopy},}\ }\href {https://doi.org/10.1021/acs.jpca.6b06060} {\bibfield
   {journal} {\bibinfo  {journal} {J. Phys. Chem. A}\ }\textbf {\bibinfo
  {volume} {120}},\ \bibinfo {pages} {6745--6752} (\bibinfo {year}
  {2016})}\BibitemShut {NoStop}%
\bibitem [{\citenamefont {Nan}, \citenamefont {Burak}, and\ \citenamefont
  {Houston}(1993)}]{Nan:CPL209:383}%
  \BibitemOpen
  \bibfield  {author} {\bibinfo {author} {\bibfnamefont {G.}~\bibnamefont
  {Nan}}, \bibinfo {author} {\bibfnamefont {I.}~\bibnamefont {Burak}}, and\
  \bibinfo {author} {\bibfnamefont {P.}~\bibnamefont {Houston}},\ }\bibfield
  {title} {\enquote {\bibinfo {title} {Photodissociation of {OCS} at 222 nm.
  {T}he triplet channel},}\ }\href
  {https://doi.org/https://doi.org/10.1016/0009-2614(93)80035-N} {\bibfield
  {journal} {\bibinfo  {journal} {Chem. Phys. Lett.}\ }\textbf {\bibinfo
  {volume} {209}},\ \bibinfo {pages} {383 -- 389} (\bibinfo {year}
  {1993})}\BibitemShut {NoStop}%
\bibitem [{\citenamefont {Mo}\ \emph {et~al.}(1996)\citenamefont {Mo},
  \citenamefont {Katayanagi}, \citenamefont {Heaven}, and\ \citenamefont
  {Suzuki}}]{Mo:PRL77:830}%
  \BibitemOpen
  \bibfield  {author} {\bibinfo {author} {\bibfnamefont {Y.}~\bibnamefont
  {Mo}}, \bibinfo {author} {\bibfnamefont {H.}~\bibnamefont {Katayanagi}},
  \bibinfo {author} {\bibfnamefont {M.~C.}\ \bibnamefont {Heaven}}, and\
  \bibinfo {author} {\bibfnamefont {T.}~\bibnamefont {Suzuki}},\ }\bibfield
  {title} {\enquote {\bibinfo {title} {Simultaneous measurement of recoil
  velocity and alignment of {$S(^{1}D_{2})$} atoms in photodissociation of
  {OCS}},}\ }\href {https://doi.org/10.1103/PhysRevLett.77.830} {\bibfield
  {journal} {\bibinfo  {journal} {Phys. Rev. Lett.}\ }\textbf {\bibinfo
  {volume} {77}},\ \bibinfo {pages} {830--833} (\bibinfo {year}
  {1996})}\BibitemShut {NoStop}%
\bibitem [{\citenamefont {Rakitzis}, \citenamefont {Samartzis}, and\
  \citenamefont {Kitsopoulos}(2001)}]{Rakitzis:PRL87:123001}%
  \BibitemOpen
  \bibfield  {author} {\bibinfo {author} {\bibfnamefont {T.~P.}\ \bibnamefont
  {Rakitzis}}, \bibinfo {author} {\bibfnamefont {P.~C.}\ \bibnamefont
  {Samartzis}}, and\ \bibinfo {author} {\bibfnamefont {T.~N.}\ \bibnamefont
  {Kitsopoulos}},\ }\bibfield  {title} {\enquote {\bibinfo {title} {Complete
  measurement of {S$(^1\!D_2)$} photofragment alignment from {A}bel-invertible
  ion images},}\ }\href {https://doi.org/10.1103/PhysRevLett.87.123001}
  {\bibfield  {journal} {\bibinfo  {journal} {Phys. Rev. Lett.}\ }\textbf
  {\bibinfo {volume} {87}},\ \bibinfo {pages} {123001} (\bibinfo {year}
  {2001})}\BibitemShut {NoStop}%
\bibitem [{\citenamefont {Lee}\ \emph {et~al.}(2006)\citenamefont {Lee},
  \citenamefont {Silva}, \citenamefont {Thamanna}, \citenamefont
  {Vasyutinskii}, and\ \citenamefont {Suits}}]{Lee:JCP125:144318}%
  \BibitemOpen
  \bibfield  {author} {\bibinfo {author} {\bibfnamefont {S.~K.}\ \bibnamefont
  {Lee}}, \bibinfo {author} {\bibfnamefont {R.}~\bibnamefont {Silva}}, \bibinfo
  {author} {\bibfnamefont {S.}~\bibnamefont {Thamanna}}, \bibinfo {author}
  {\bibfnamefont {O.~S.}\ \bibnamefont {Vasyutinskii}}, and\ \bibinfo {author}
  {\bibfnamefont {A.~G.}\ \bibnamefont {Suits}},\ }\bibfield  {title} {\enquote
  {\bibinfo {title} {{S($^1\!D_{2}$)} atomic orbital polarization in the
  photodissociation of {OCS} at 193 nm: {C}onstruction of the complete density
  matrix},}\ }\href {https://doi.org/10.1063/1.2357948} {\bibfield  {journal}
  {\bibinfo  {journal} {J. Chem. Phys.}\ }\textbf {\bibinfo {volume} {125}},\
  \bibinfo {pages} {144318} (\bibinfo {year} {2006})}\BibitemShut {NoStop}%
\bibitem [{\citenamefont {Schmidt} and\ \citenamefont
  {Olsen}(2014)}]{Schmidt:JCP141:184310}%
  \BibitemOpen
  \bibfield  {author} {\bibinfo {author} {\bibfnamefont {J.~A.}\ \bibnamefont
  {Schmidt}} and\ \bibinfo {author} {\bibfnamefont {J.~M.~H.}\ \bibnamefont
  {Olsen}},\ }\bibfield  {title} {\enquote {\bibinfo {title} {Photodissociation
  of {OCS}: {D}eviations between theory and experiment, and the importance of
  higher order correlation effects},}\ }\href
  {https://doi.org/10.1063/1.4901426} {\bibfield  {journal} {\bibinfo
  {journal} {J. Chem. Phys.}\ }\textbf {\bibinfo {volume} {141}},\ \bibinfo
  {pages} {184310} (\bibinfo {year} {2014})}\BibitemShut {NoStop}%
\bibitem [{\citenamefont {Karamatskos}\ \emph
  {et~al.}(2019{\natexlab{b}})\citenamefont {Karamatskos}, \citenamefont
  {Raabe}, \citenamefont {Mullins}, \citenamefont {Trabattoni}, \citenamefont
  {Stammer}, \citenamefont {Goldsztejn}, \citenamefont {Johansen},
  \citenamefont {D{\l}ugo{\l}\k{e}cki}, \citenamefont {Stapelfeldt},
  \citenamefont {Vrakking}, \citenamefont {Trippel}, \citenamefont {Rouzée},\
  and\ \citenamefont {Küpper}}]{Karamatskos:NatComm10:3364}%
  \BibitemOpen
  \bibfield  {author} {\bibinfo {author} {\bibfnamefont {E.~T.}\ \bibnamefont
  {Karamatskos}}, \bibinfo {author} {\bibfnamefont {S.}~\bibnamefont {Raabe}},
  \bibinfo {author} {\bibfnamefont {T.}~\bibnamefont {Mullins}}, \bibinfo
  {author} {\bibfnamefont {A.}~\bibnamefont {Trabattoni}}, \bibinfo {author}
  {\bibfnamefont {P.}~\bibnamefont {Stammer}}, \bibinfo {author} {\bibfnamefont
  {G.}~\bibnamefont {Goldsztejn}}, \bibinfo {author} {\bibfnamefont {R.~R.}\
  \bibnamefont {Johansen}}, \bibinfo {author} {\bibfnamefont {K.}~\bibnamefont
  {D{\l}ugo{\l}\k{e}cki}}, \bibinfo {author} {\bibfnamefont {H.}~\bibnamefont
  {Stapelfeldt}}, \bibinfo {author} {\bibfnamefont {M.~J.~J.}\ \bibnamefont
  {Vrakking}}, \bibinfo {author} {\bibfnamefont {S.}~\bibnamefont {Trippel}},
  \bibinfo {author} {\bibfnamefont {A.}~\bibnamefont {Rouzée}}, and\ \bibinfo
  {author} {\bibfnamefont {J.}~\bibnamefont {Küpper}},\ }\bibfield  {title}
  {\enquote {\bibinfo {title} {Molecular movie of ultrafast coherent rotational
  dynamics of {OCS}},}\ }\href {https://doi.org/10.1038/s41467-019-11122-y}
  {\bibfield  {journal} {\bibinfo  {journal} {Nat. Commun.}\ }\textbf {\bibinfo
  {volume} {10}},\ \bibinfo {pages} {3364} (\bibinfo {year}
  {2019}{\natexlab{b}})},\ \Eprint {https://arxiv.org/abs/1807.01034}
  {arXiv:1807.01034 [physics]}\BibitemShut {NoStop}%
\bibitem [{\citenamefont {Wiese}\ \emph {et~al.}(2020)\citenamefont {Wiese},
  \citenamefont {Onvlee}, \citenamefont {Trippel}, and\ \citenamefont
  {Küpper}}]{Wiese:PRR2020:inprep}%
  \BibitemOpen
  \bibfield  {author} {\bibinfo {author} {\bibfnamefont {J.}~\bibnamefont
  {Wiese}}, \bibinfo {author} {\bibfnamefont {J.}~\bibnamefont {Onvlee}},
  \bibinfo {author} {\bibfnamefont {S.}~\bibnamefont {Trippel}}, and\ \bibinfo
  {author} {\bibfnamefont {J.}~\bibnamefont {Küpper}},\ }\href@noop {}
  {\enquote {\bibinfo {title} {Strong-field ionization of complex molecules},}\
  } (\bibinfo {year} {2020}),\ \bibinfo {note} {under review},\ \Eprint
  {https://arxiv.org/abs/2003.02116} {arXiv:2003.02116 [physics]}\BibitemShut
  {NoStop}%
\bibitem [{\citenamefont {Hillenkamp}, \citenamefont {Keinan}, and\
  \citenamefont {Even}(2003)}]{Hillenkamp:JCP118:8699}%
  \BibitemOpen
  \bibfield  {author} {\bibinfo {author} {\bibfnamefont {M.}~\bibnamefont
  {Hillenkamp}}, \bibinfo {author} {\bibfnamefont {S.}~\bibnamefont {Keinan}},\
  and\ \bibinfo {author} {\bibfnamefont {U.}~\bibnamefont {Even}},\ }\bibfield
  {title} {\enquote {\bibinfo {title} {Condensation limited cooling in
  supersonic expansions},}\ }\href {https://doi.org/10.1063/1.1568331}
  {\bibfield  {journal} {\bibinfo  {journal} {J. Chem. Phys.}\ }\textbf
  {\bibinfo {volume} {118}},\ \bibinfo {pages} {8699--8705} (\bibinfo {year}
  {2003})}\BibitemShut {NoStop}%
\bibitem [{\citenamefont {Dribinski}\ \emph {et~al.}(2002)\citenamefont
  {Dribinski}, \citenamefont {Ossadtchi}, \citenamefont {Mandelshtam}, and\
  \citenamefont {Reisler}}]{Dribinski:RSI73:2634}%
  \BibitemOpen
  \bibfield  {author} {\bibinfo {author} {\bibfnamefont {V.}~\bibnamefont
  {Dribinski}}, \bibinfo {author} {\bibfnamefont {A.}~\bibnamefont
  {Ossadtchi}}, \bibinfo {author} {\bibfnamefont {V.~A.}\ \bibnamefont
  {Mandelshtam}}, and\ \bibinfo {author} {\bibfnamefont {H.}~\bibnamefont
  {Reisler}},\ }\bibfield  {title} {\enquote {\bibinfo {title} {{Reconstruction
  of {A}bel-transformable images: {T}he {G}aussian basis-set expansion {A}bel
  transform method}},}\ }\href {https://doi.org/10.1063/1.1482156} {\bibfield
  {journal} {\bibinfo  {journal} {Rev. Sci. Instrum.}\ }\textbf {\bibinfo
  {volume} {73}},\ \bibinfo {pages} {2634} (\bibinfo {year}
  {2002})}\BibitemShut {NoStop}%
\bibitem [{\citenamefont {Meyer}, \citenamefont {Manthe}, and\ \citenamefont
  {Cederbaum}(1990)}]{Meyer:CPL165:73}%
  \BibitemOpen
  \bibfield  {author} {\bibinfo {author} {\bibfnamefont {H.-D.}\ \bibnamefont
  {Meyer}}, \bibinfo {author} {\bibfnamefont {U.}~\bibnamefont {Manthe}}, and\
  \bibinfo {author} {\bibfnamefont {L.}~\bibnamefont {Cederbaum}},\ }\bibfield
  {title} {\enquote {\bibinfo {title} {The multi-configurational time-dependent
  {H}artree approach},}\ }\href {https://doi.org/10.1016/0009-2614(90)87014-I}
  {\bibfield  {journal} {\bibinfo  {journal} {Chem. Phys. Lett.}\ }\textbf
  {\bibinfo {volume} {165}},\ \bibinfo {pages} {73--78} (\bibinfo {year}
  {1990})}\BibitemShut {NoStop}%
\bibitem [{\citenamefont {Manthe}, \citenamefont {Meyer}, and\ \citenamefont
  {Cederbaum}(1992)}]{Manthe:JCP97:3199}%
  \BibitemOpen
  \bibfield  {author} {\bibinfo {author} {\bibfnamefont {U.}~\bibnamefont
  {Manthe}}, \bibinfo {author} {\bibfnamefont {H.~D.}\ \bibnamefont {Meyer}},\
  and\ \bibinfo {author} {\bibfnamefont {L.~S.}\ \bibnamefont {Cederbaum}},\
  }\bibfield  {title} {\enquote {\bibinfo {title} {Wave‐packet dynamics
  within the multiconfiguration {H}artree framework: {G}eneral aspects and
  application to {NOCl}},}\ }\href {https://doi.org/10.1063/1.463007}
  {\bibfield  {journal} {\bibinfo  {journal} {J. Phys. Chem.}\ }\textbf
  {\bibinfo {volume} {97}},\ \bibinfo {pages} {3199--3213} (\bibinfo {year}
  {1992})}\BibitemShut {NoStop}%
\bibitem [{\citenamefont {Welsch}, \citenamefont {Huarte-Larrañaga}, and\
  \citenamefont {Manthe}(2012)}]{Welsch:JCP136:064117}%
  \BibitemOpen
  \bibfield  {author} {\bibinfo {author} {\bibfnamefont {R.}~\bibnamefont
  {Welsch}}, \bibinfo {author} {\bibfnamefont {F.}~\bibnamefont
  {Huarte-Larrañaga}}, and\ \bibinfo {author} {\bibfnamefont
  {U.}~\bibnamefont {Manthe}},\ }\bibfield  {title} {\enquote {\bibinfo {title}
  {State-to-state reaction probabilities within the quantum transition state
  framework},}\ }\href {https://doi.org/10.1063/1.3684631} {\bibfield
  {journal} {\bibinfo  {journal} {J. Chem. Phys.}\ }\textbf {\bibinfo {volume}
  {136}},\ \bibinfo {pages} {064117} (\bibinfo {year} {2012})}\BibitemShut
  {NoStop}%
\bibitem [{\citenamefont {Welsch} and\ \citenamefont
  {Manthe}(2012)}]{Welsch:MP110:703}%
  \BibitemOpen
  \bibfield  {author} {\bibinfo {author} {\bibfnamefont {R.}~\bibnamefont
  {Welsch}} and\ \bibinfo {author} {\bibfnamefont {U.}~\bibnamefont
  {Manthe}},\ }\bibfield  {title} {\enquote {\bibinfo {title} {Thermal flux
  based analysis of state-to-state reaction probabilities},}\ }\href
  {https://doi.org/10.1080/00268976.2012.657803} {\bibfield  {journal}
  {\bibinfo  {journal} {Mol. Phys.}\ }\textbf {\bibinfo {volume} {110}},\
  \bibinfo {pages} {703--715} (\bibinfo {year} {2012})}\BibitemShut {NoStop}%
\bibitem [{\citenamefont {Manthe}(1996)}]{Manthe:JCP105:6989}%
  \BibitemOpen
  \bibfield  {author} {\bibinfo {author} {\bibfnamefont {U.}~\bibnamefont
  {Manthe}},\ }\bibfield  {title} {\enquote {\bibinfo {title} {A
  time‐dependent discrete variable representation for (multiconfiguration)
  {H}artree methods},}\ }\href {https://doi.org/10.1063/1.471847} {\bibfield
  {journal} {\bibinfo  {journal} {J. Chem. Phys.}\ }\textbf {\bibinfo {volume}
  {105}},\ \bibinfo {pages} {6989--6994} (\bibinfo {year} {1996})}\BibitemShut
  {NoStop}%
\bibitem [{\citenamefont {Manthe}(2006)}]{Manthe:CP329:168}%
  \BibitemOpen
  \bibfield  {author} {\bibinfo {author} {\bibfnamefont {U.}~\bibnamefont
  {Manthe}},\ }\bibfield  {title} {\enquote {\bibinfo {title} {On the
  integration of the multi-configurational time-dependent {H}artree ({MCTDH})
  equations of motion},}\ }\href
  {https://doi.org/10.1016/j.chemphys.2006.05.028} {\bibfield  {journal}
  {\bibinfo  {journal} {Chem. Phys.}\ }\textbf {\bibinfo {volume} {329}},\
  \bibinfo {pages} {168--178} (\bibinfo {year} {2006})}\BibitemShut {NoStop}%
\bibitem [{\citenamefont {Morse}\ \emph {et~al.}(1999)\citenamefont {Morse},
  \citenamefont {Takahashi}, \citenamefont {Eland}, and\ \citenamefont
  {Karlsson}}]{Morse:IJMSIP184:67}%
  \BibitemOpen
  \bibfield  {author} {\bibinfo {author} {\bibfnamefont {S.}~\bibnamefont
  {Morse}}, \bibinfo {author} {\bibfnamefont {M.}~\bibnamefont {Takahashi}},
  \bibinfo {author} {\bibfnamefont {J.~H.~D.}\ \bibnamefont {Eland}}, and\
  \bibinfo {author} {\bibfnamefont {L.}~\bibnamefont {Karlsson}},\ }\bibfield
  {title} {\enquote {\bibinfo {title} {Dissociative photoionisation of {OCS}
  from threshold to 40.8~{eV}},}\ }\href
  {https://doi.org/https://doi.org/10.1016/S1387-3806(98)14254-9} {\bibfield
  {journal} {\bibinfo  {journal} {Int. J. Mass Spectrom. Ion Processes}\
  }\textbf {\bibinfo {volume} {184}},\ \bibinfo {pages} {67--74} (\bibinfo
  {year} {1999})}\BibitemShut {NoStop}%
\bibitem [{\citenamefont {Stapelfeldt}\ \emph {et~al.}(1998)\citenamefont
  {Stapelfeldt}, \citenamefont {Constant}, \citenamefont {Sakai}, and\
  \citenamefont {Corkum}}]{Stapelfeldt:PRA58:426}%
  \BibitemOpen
  \bibfield  {author} {\bibinfo {author} {\bibfnamefont {H.}~\bibnamefont
  {Stapelfeldt}}, \bibinfo {author} {\bibfnamefont {E.}~\bibnamefont
  {Constant}}, \bibinfo {author} {\bibfnamefont {H.}~\bibnamefont {Sakai}},\
  and\ \bibinfo {author} {\bibfnamefont {P.~B.}\ \bibnamefont {Corkum}},\
  }\bibfield  {title} {\enquote {\bibinfo {title} {Time-resolved {C}oulomb
  explosion imaging: A method to measure structure and dynamics of molecular
  nuclear wave packets},}\ }\href {https://doi.org/10.1103/PhysRevA.58.426}
  {\bibfield  {journal} {\bibinfo  {journal} {Phys. Rev. A}\ }\textbf {\bibinfo
  {volume} {58}},\ \bibinfo {pages} {426--433} (\bibinfo {year}
  {1998})}\BibitemShut {NoStop}%
\bibitem [{\citenamefont {Ammosov}, \citenamefont {Delone}, and\ \citenamefont
  {Krainov}(1986)}]{Ammosov:SVJETP64:1191}%
  \BibitemOpen
  \bibfield  {author} {\bibinfo {author} {\bibfnamefont {M.~V.}\ \bibnamefont
  {Ammosov}}, \bibinfo {author} {\bibfnamefont {N.~B.}\ \bibnamefont
  {Delone}}, and\ \bibinfo {author} {\bibfnamefont {V.~P.}\ \bibnamefont
  {Krainov}},\ }\bibfield  {title} {\enquote {\bibinfo {title} {Tunnel
  ionization of complex atoms and of atomic ions in an alternating
  electromagnetic field},}\ }\href
  {http://www.jetp.ac.ru/cgi-bin/e/index/e/64/6/p1191?a=list} {\bibfield
  {journal} {\bibinfo  {journal} {Soviet Physics - JETP}\ }\textbf {\bibinfo
  {volume} {64}},\ \bibinfo {pages} {1191--1194} (\bibinfo {year}
  {1986})}\BibitemShut {NoStop}%
\bibitem [{\citenamefont {Wilkinson}\ \emph {et~al.}(2014)\citenamefont
  {Wilkinson}, \citenamefont {Boguslavskiy}, \citenamefont {Mikosch},
  \citenamefont {Bertrand}, \citenamefont {Wörner}, \citenamefont
  {Villeneuve}, \citenamefont {Spanner}, \citenamefont {Patchkovskii}, and\
  \citenamefont {Stolow}}]{Wilkinson:JCP140:204301}%
  \BibitemOpen
  \bibfield  {author} {\bibinfo {author} {\bibfnamefont {I.}~\bibnamefont
  {Wilkinson}}, \bibinfo {author} {\bibfnamefont {A.~E.}\ \bibnamefont
  {Boguslavskiy}}, \bibinfo {author} {\bibfnamefont {J.}~\bibnamefont
  {Mikosch}}, \bibinfo {author} {\bibfnamefont {J.~B.}\ \bibnamefont
  {Bertrand}}, \bibinfo {author} {\bibfnamefont {H.~J.}\ \bibnamefont
  {Wörner}}, \bibinfo {author} {\bibfnamefont {D.~M.}\ \bibnamefont
  {Villeneuve}}, \bibinfo {author} {\bibfnamefont {M.}~\bibnamefont {Spanner}},
  \bibinfo {author} {\bibfnamefont {S.}~\bibnamefont {Patchkovskii}}, and\
  \bibinfo {author} {\bibfnamefont {A.}~\bibnamefont {Stolow}},\ }\bibfield
  {title} {\enquote {\bibinfo {title} {Excited state dynamics in {SO}$_2$. i.
  bound state relaxation studied by time-resolved photoelectron-photoion
  coincidence spectroscopy},}\ }\href {https://doi.org/10.1063/1.4875035}
  {\bibfield  {journal} {\bibinfo  {journal} {J. Chem. Phys.}\ }\textbf
  {\bibinfo {volume} {140}},\ \bibinfo {pages} {204301} (\bibinfo {year}
  {2014})}\BibitemShut {NoStop}%
\bibitem [{\citenamefont {Mai}, \citenamefont {Marquetand}, and\ \citenamefont
  {González}(2014)}]{Mai:JCP140:204302}%
  \BibitemOpen
  \bibfield  {author} {\bibinfo {author} {\bibfnamefont {S.}~\bibnamefont
  {Mai}}, \bibinfo {author} {\bibfnamefont {P.}~\bibnamefont {Marquetand}},\
  and\ \bibinfo {author} {\bibfnamefont {L.}~\bibnamefont {González}},\
  }\bibfield  {title} {\enquote {\bibinfo {title} {Non-adiabatic and
  intersystem crossing dynamics in {SO}$_2$. ii. the role of triplet states in
  the bound state dynamics studied by surface-hopping simulations},}\ }\href
  {https://doi.org/10.1063/1.4875036} {\bibfield  {journal} {\bibinfo
  {journal} {J. Chem. Phys.}\ }\textbf {\bibinfo {volume} {140}},\ \bibinfo
  {pages} {204302} (\bibinfo {year} {2014})}\BibitemShut {NoStop}%
\bibitem [{\citenamefont {Owens}\ \emph {et~al.}(2018)\citenamefont {Owens},
  \citenamefont {Yachmenev}, \citenamefont {Yurchenko}, and\ \citenamefont
  {K\"{u}pper}}]{Owens:PRL121:193201}%
  \BibitemOpen
  \bibfield  {author} {\bibinfo {author} {\bibfnamefont {A.}~\bibnamefont
  {Owens}}, \bibinfo {author} {\bibfnamefont {A.}~\bibnamefont {Yachmenev}},
  \bibinfo {author} {\bibfnamefont {S.~N.}\ \bibnamefont {Yurchenko}}, and\
  \bibinfo {author} {\bibfnamefont {J.}~\bibnamefont {K\"{u}pper}},\ }\bibfield
   {title} {\enquote {\bibinfo {title} {{C}limbing the {R}otational {L}adder to
  {C}hirality},}\ }\href {https://doi.org/10.1103/physrevlett.121.193201}
  {\bibfield  {journal} {\bibinfo  {journal} {Phys. Rev. Lett.}\ }\textbf
  {\bibinfo {volume} {121}},\ \bibinfo {pages} {193201} (\bibinfo {year}
  {2018})},\ \Eprint {https://arxiv.org/abs/1802.07803} {arXiv:1802.07803
  [physics]}\BibitemShut {NoStop}%
\bibitem [{\citenamefont {Pathak}\ \emph {et~al.}(2020)\citenamefont {Pathak},
  \citenamefont {Ibele}, \citenamefont {Boll}, \citenamefont {Callegari},
  \citenamefont {Demidovich}, \citenamefont {Erk}, \citenamefont {Feifel},
  \citenamefont {Forbes}, \citenamefont {Di~Fraia}, \citenamefont {Giannessi},
  \citenamefont {Hansen}, \citenamefont {Holland}, \citenamefont {Ingle},
  \citenamefont {Mason}, \citenamefont {Plekan}, \citenamefont {Prince},
  \citenamefont {Rouzee}, \citenamefont {Squibb}, \citenamefont {Tross},
  \citenamefont {Ashfold}, \citenamefont {Curchod}, and\ \citenamefont
  {Rolles}}]{Pathak:NatChem12:795}%
  \BibitemOpen
  \bibfield  {author} {\bibinfo {author} {\bibfnamefont {S.}~\bibnamefont
  {Pathak}}, \bibinfo {author} {\bibfnamefont {L.~M.}\ \bibnamefont {Ibele}},
  \bibinfo {author} {\bibfnamefont {R.}~\bibnamefont {Boll}}, \bibinfo {author}
  {\bibfnamefont {C.}~\bibnamefont {Callegari}}, \bibinfo {author}
  {\bibfnamefont {A.}~\bibnamefont {Demidovich}}, \bibinfo {author}
  {\bibfnamefont {B.}~\bibnamefont {Erk}}, \bibinfo {author} {\bibfnamefont
  {R.}~\bibnamefont {Feifel}}, \bibinfo {author} {\bibfnamefont
  {R.}~\bibnamefont {Forbes}}, \bibinfo {author} {\bibfnamefont
  {M.}~\bibnamefont {Di~Fraia}}, \bibinfo {author} {\bibfnamefont
  {L.}~\bibnamefont {Giannessi}}, \bibinfo {author} {\bibfnamefont {C.~S.}\
  \bibnamefont {Hansen}}, \bibinfo {author} {\bibfnamefont {D.~M.~P.}\
  \bibnamefont {Holland}}, \bibinfo {author} {\bibfnamefont {R.~A.}\
  \bibnamefont {Ingle}}, \bibinfo {author} {\bibfnamefont {R.}~\bibnamefont
  {Mason}}, \bibinfo {author} {\bibfnamefont {O.}~\bibnamefont {Plekan}},
  \bibinfo {author} {\bibfnamefont {K.~C.}\ \bibnamefont {Prince}}, \bibinfo
  {author} {\bibfnamefont {A.}~\bibnamefont {Rouzee}}, \bibinfo {author}
  {\bibfnamefont {R.~J.}\ \bibnamefont {Squibb}}, \bibinfo {author}
  {\bibfnamefont {J.}~\bibnamefont {Tross}}, \bibinfo {author} {\bibfnamefont
  {M.~N.~R.}\ \bibnamefont {Ashfold}}, \bibinfo {author} {\bibfnamefont
  {B.~F.~E.}\ \bibnamefont {Curchod}}, and\ \bibinfo {author} {\bibfnamefont
  {D.}~\bibnamefont {Rolles}},\ }\bibfield  {title} {\enquote {\bibinfo {title}
  {Tracking the ultraviolet-induced photochemistry of thiophenone during and
  after ultrafast ring opening},}\ }\href
  {https://doi.org/10.1038/s41557-020-0507-3} {\bibfield  {journal} {\bibinfo
  {journal} {Nat. Chem.}\ }\textbf {\bibinfo {volume} {12}},\ \bibinfo {pages}
  {795} (\bibinfo {year} {2020})}\BibitemShut {NoStop}%
\end{thebibliography}%
\onecolumngrid%
\listofnotes%
\end{document}